\documentclass[structabstract]{aa}  
\newcommand{\nc}{\newcommand}
\nc{\cm}{\,\mathrm{cm}}
\nc{\micron}{\,\mathrm{\mu m}}
\nc{\cmcm}{\,\mathrm{cm^{-2}}}
\nc{\cmcmg}{\,\mathrm{cm^{2}\,g^{-1}}}
\nc{\iccm}{\,\mathrm{cm^{-3}}}
\nc{\kmps}{\,\mathrm{km\, s^{-1}}}       
\nc{\Kkmps}{\,\mathrm{K\, km\, s^{-1}}}       
\nc{\K}{\,\mathrm{K}}       
\nc{\vlsr}{v_\mathrm{LSR}}       
\nc{\etamb}{\eta_\mathrm{mb}}       
\nc{\vt}{v_\mathrm{turb}}       
\nc{\dv}{\Delta v}       
\nc{\trot}{T_\mathrm{rot}}      
\nc{\tkin}{T_\mathrm{kin}}      
\nc{\tex}{T_\mathrm{ex}}      
\nc{\nhh}{n(\mathrm{H_2})}      
\nc{\Nhh}{N(\mathrm{H_2})}      
\nc{\nmol}{N_\mathrm{mol}}      
\nc{\Xmol}[1]{$X[\mathrm{#1}]$} 
\nc{\ebf}{\eta_\mathrm{bf}}      
\nc{\jkk}[4]{${#1}_{#2}-{#3}_{#4}$}
\nc{\me}[2]{#1\times 10^{#2}}
\nc{\hh}{$\mathrm{H_2}$}
\nc{\hhco}{$\mathrm{H_2CO}$}
\nc{\hhico}{$\mathrm{H_2^{13}CO}$}
\nc{\hdco}{$\mathrm{HDCO}$}
\nc{\ddco}{$\mathrm{D_2CO}$}
\nc{\chhhoh}{$\mathrm{CH_3OH}$}
\nc{\nhhh}{$\mathrm{NH_3}$}
\nc{\oio}{$\mathrm{O^{18}O}$}
\nc{\co}{$\mathrm{CO}$}
\nc{\cio}{$\mathrm{C^{18}O}$}
\nc{\icio}{$\mathrm{^{13}C^{18}O}$}
\nc{\ico}{$\mathrm{^{13}CO}$}
\nc{\so}{$\mathrm{SO}$}
\nc{\iso}{$\mathrm{^{34}SO}$}
\nc{\isoo}{$\mathrm{^{34}SO_2}$}
\nc{\soo}{$\mathrm{SO_2}$}
\nc{\nndp}{$\mathrm{N_2D^+}$}
\nc{\nnhp}{$\mathrm{N_2H^+}$}
\nc{\ccchh}{$c-\mathrm{C_3H_2}$}
\nc{\rhooph}{$\rho$~Ophiuchi}
\nc{\roa}{$\rho$~Oph~A}
\nc{\roc}{$\rho$~Oph~C}
\nc{\mc}[1]{\multicolumn{1}{c}{#1}}
\nc{\sref}[1]{Sect.~(\ref{#1})}
\nc{\fref}[1]{Fig.~\ref{#1}}
\nc{\frefs}[2]{Figs.~\ref{#1} and \ref{#2}}
\nc{\tref}[1]{Table~\ref{#1}}
\nc{\eref}[1]{Eq.~(\ref{#1})}
\nc{\RA}[4]{$\alpha(\mathrm{#1})={#2}^\mathrm{h}\,{#3}^\mathrm{m}\,{#4}^\mathrm{s}$}
\nc{\Dec}[4]{$\delta(\mathrm{#1})={#2}\degr\,{#3}\arcmin\,{#4}\arcsec$}
\usepackage{graphicx}
\usepackage{txfonts}
\usepackage{natbib}
\bibpunct{(}{)}{;}{a}{}{,} 
\begin{document}
   \title{Deuterated formaldehyde in $\rho$ Ophiuchi A
   \thanks{Based on observations with the Atacama Pathfinder EXperiment (APEX)
   telescope. APEX is a collaboration between the Max-Planck-Institut f{\"u}r
   Radioastronomie, the European Southern Observatory, and the Onsala Space
   Observatory}
   }


   \author{P. Bergman\inst{1}
        \and
           B. Parise\inst{2}
	\and
	   R. Liseau\inst{3}
	\and
	   B. Larsson\inst{4}
        }

   \institute{Onsala Space Observatory, Chalmers University of Technology,
              SE-439 92 Onsala, Sweden
              \email{per.bergman@chalmers.se}
         \and
              Max Planck Institut f{\"u}r Radioastronomie, Auf dem H{\"u}gel 69,
	      53121 Bonn, Germany
	 \and
	      Department of Earth and Space Sciences, Chalmers University of Technology,
              SE-439 92 Onsala, Sweden\ 
	 \and
	      Department of Astronomy, Stockholm University, AlbaNova,
	      SE-10691 Stockholm, Sweden
             }

   \date{Received ? ; accepted ?}

 
  \abstract
   {Formaldehyde is an organic molecule that is abundant in the interstellar
   medium. High deuterium fractionation is a common feature in low-mass
   star-forming regions. Observing several isotopologues of molecules
   is an excellent tool for understanding the formation paths of the molecules.}
   {We seek an understanding of how the various deuterated isotopologues
   of formaldehyde are formed in the dense regions of low-mass star formation.
   More specifically, we adress the question of how the very high
   deuteration levels (several orders of magnitude above the cosmic
   D/H ratio) can occur using {\hhco} data of the nearby {\roa} molecular
   cloud.}
   {From mapping observations of {\hhco}, {\hdco}, and {\ddco},
   we have determined how the degree of
   deuterium fractionation changes over the central $3'\times 3'$ region
   of {\roa}. The multi-transition data of the various
   {\hhco} isotopologues, as well as from other molecules (e.g., {\chhhoh} and
   {\nndp})
   present in the observed bands,
   were analysed using both the standard type rotation diagram analysis
   and, in selected cases, a more elaborate method of solving
   the radiative transfer for optically thick emission. In addition to molecular
   column densities, the analysis also
   estimates the kinetic temperature and {\hh} density.}
   {Toward the SM1 core in {\roa}, the {\hhco} deuterium
   fractionation is very high. In fact, the observed {\ddco}/{\hdco} ratio is
   $1.34\pm 0.19$, while the {\hdco}/{\hhco} ratio is $0.107\pm 0.015$. This
   is the first time, to our knowledge, that the {\ddco}/{\hdco} abundance ratio is observed to
   be greater than 1. The kinetic temperature is in the range 20-30 K
   in the cores of {\roa}, and the {\hh} density is $\me{(6-10)}{5}\iccm$.
   We estimate that the total {\hh} column density toward the deuterium peak
   is $\me{(1-4)}{23}\cmcm$.
   As depleted gas-phase chemistry is not adequate, we suggest that grain
   chemistry, possibly due to abstraction and exchange reactions along the
   reaction chain $\mathrm{H_2CO \to HDCO \to D_2CO}$, is at work to produce
   the very high deuterium levels observed.}
   {}

   \keywords{astrochemistry --
             ISM: abundances --
	     ISM: clouds --
	     ISM: individual objects: \object{$\rho$ Oph A} --
             ISM: molecules
            }

   \maketitle
%

\section{Introduction}

   The study of deuterated molecules in the interstellar medium (ISM) has
   been intensified over the past decade ever since it was discovered
   that singly and multiply deuterated species (like the D-containing
   versions of {\hhco}, {\chhhoh}, {\nhhh}) occurred at abundances
   that were orders of magnitude higher than would be expected from the
   local ISM D/H
   ratio of about $\me{1.5}{-5}$ \citep{2003SSRv..106...49L}. The regions
   that show these elevated D-abundances are mainly associated with
   low-mass protostars. For instance, around IRAS16293$-$2422,
   \citet{1998A&A...338L..43C} and later \citet{2000A&A...359.1169L} detected lines
   from doubly-deuterated
   formaldehyde ({\ddco}) which previously had only been seen in the Orion~KL
   ridge \citep{1990ApJ...362L..29T}. Later, \citet{2002P&SS...50.1205L} extended the
   study to a larger set of protostars arriving at {\ddco}/{\hhco} abundance
   ratios as high as 0.05-0.4. Likewise, high degrees of deuterium
   fractionation in {\chhhoh} \citep{2002A&A...393L..49P, 2004A&A...416..159P,
   2006A&A...453..949P} and {\nhhh} \citep{2000A&A...354L..63R, 2001ApJ...552L.163L,
   2002ApJ...571L..55L} were subsequently discovered.
   
   These studies are
   all related to single telescope pointings. To our knowledge, only one
   effort to delineate the {\ddco}/{\hhco} abundance variation within a
   single source (IRAS16293$-$2422) has been made \citep{2001A&A...372..998C}. The
   {\ddco}/{\hhco} ratio peaks some $20''$ away from the protostar. Further out, the
   degree of deuteration seems to be lower in the quiscent gas \citep{2002A&A...381L..17C}.
   In the case of deuterated ammonia, the mapping study by \citet{2005A&A...438..585R}
   revealed that the deuterium peak is typically offset from the positions of the
   embedded protostars. These authors argued that the observed scenario could be explained
   by the formation of deuterium-enriched ices during the cold pre-collapse phase.
   At a later stage, a newly formed protostar evaporates the ices.
   
   In their
   extensive study of deuterated {\hhco} and {\chhhoh}, \citet{2006A&A...453..949P}
   concluded that formation of {\chhhoh} on grain surfaces was a likely
   explanation for the high degrees of deuterium fractionation seen in the various
   isotopologues. For {\hhco}, the
   situation was less clear and a formation path involving gas-phase reactions
   could not be ruled out. In fact, for the warmer Orion Bar PDR region,
   \citet{2009A&A...508..737P} very recently advocated that gas-phase
   chemistry is entirely responsible for the deuterium
   fractionation seen for some singly deuterated species (including {\hdco}).
   It should be noted that singly-deuterated species have been detected in
   dark and translucent clouds \citep{2001ApJS..136..579T} where the deuteration
   occurs only in the dense parts.
   
   In interstellar clouds, deuterium is mostly molecular in the form of HD.
   Deuterium can be transferred from this main molecular reservoir to other 
   molecules via exothermic reactions with the molecular
   ions $\mathrm{H_3^+}$, $\mathrm{CH_3^+}$, and $\mathrm{C_2H_2^+}$.
   These basic reactions are followed by efficient ion-neutral reactions and,
   in some cases, by reactions on grain surfaces. The exothermicity of the
   reactions involving these three molecular ions are in the range 230-550 K
   \citep{2002P&SS...50.1275G, 2004ApJ...617..685A, 1987ApJ...312..351H}.
   As a result, deuterium fractionation is efficient in  cold environments. It is
   also well established that a high degree of  depletion on the grains of CO, O, 
   and other heavy species, which would  otherwise destroy efficiently H$_3^+$ (and
   its deuterated analogues), is another important condition  for deuterium
   fractionations taking place. Correlations between CO depletion and fractionation
   have been observed towards prestellar cores by, e.g., \citet{2003ApJ...585L..55B}
   and \citet{2005ApJ...619..379C}, and in the envelope of  Class 0 protostars by
   \citet{2009A&A...493...89E}. Because of these two conditions (low temperature and
   high CO depletion), deuterium  fractionation is particularly efficient during the
   early stages of star formation. The surface reactions on cold grains important
   for deuterium fractionation of {\hhco} have recently been investigated through
   laboratory experiments \citep{2009ApJ...702..291H}.

   The aforementioned low-mass protostar IRAS16293$-$2422 is located in the
   eastern part of the {\rhooph} cloud complex. To the west in the same complex,
   more than 1 degree away, lies the {\roa} cloud \citep{1990ApJ...365..269L}
   at a distance of about 120 pc \citep{2008A&A...480..785L, 2008ApJ...675L..29L, 2008ApJ...679..512S}.
   This cloud core is well-studied by infrared, submillimeter, and millimeter continuum
   observations \citep{1989MNRAS.241..119W, 1993ApJ...406..122A, 1998A&A...336..150M}.
   It hosts a well-collimated molecular outflow \citep{1990A&A...236..180A} and its
   driving source VLA 1623 \citep{1993ApJ...406..122A}. Moreover, it was in the
   direction of this cloud core that \citet{2007A&A...466..999L}, using the
   Odin satellite, detected the 119 GHz
   line from molecular oxygen as well as the ammonia ground state line
   at 572 GHz \citep{2003A&A...402L..73L}. More recently,
   while searching for the 234 GHz {\oio} line
   \cite{2010A&A...510A..98L} detected several lines due to {\ddco} toward
   the millimeter continuum peaks. This study also includes {\cio}(3-2)
   mapping observations. Earlier, \citet{2002P&SS...50.1205L} reported
   a high {\ddco}/{\hhco} ratio toward the VLA 1623 source. The existence
   of {\ddco} in several positions of this $2'\times 3'$ cloud core formed the
   incentive of the present study as an excellent source to delineate the
   distribution of {\hhco}, {\hdco}, and {\ddco} as we know the distribution of the
   dust \citep{1998A&A...336..150M} and the gas in terms of {\cio}
   \citep{2010A&A...510A..98L}. Here we report on mapping
   observations of the {\roa} cloud core in several frequency settings that cover
   most of the low-energy {\hhco}, {\hdco}, and {\ddco} lines in the 1.3 millimeter band
   using the 12 m APEX telescope \citep{2006A&A...454L..13G}. In addition, lines from
   many other species were observed simultaneously (eg. {\chhhoh}, {\so}, and {\nndp}).
   Here the {\chhhoh} results are of importance because {\chhhoh} is
   directly involved in the {\hhco} chemistry and the
   {\nndp} results are, of course, of interest for the deuteration. Although the sulphur
   chemistry is not of immediate interest here we chose to include the {\so} and
   {\soo} observational and analysis results
   since they are of importance as a complementary tool for determining the
   physical conditions.
   This paper is organized as follows. In \sref{s:obs} we describe the observations
   and then, in \sref{s:res}, we present the observational results. In \sref{s:ana} we
   obtain the physical conditions. Before concluding, in \sref{s:con}, we discuss our
   results in \sref{s:dis}.


\section{Observations}
\label{s:obs}
   The APEX telescope at Chajnantor (Chile) was used to map the $3'\times 3'$ (with
   a spacing of $30\arcsec$) area centred on the coordinates
   \RA{J2000}{16}{26}{27.2} and \Dec{J2000}{-24}{23}{34} which is very close to the
   SM1N position in the {\roa} cloud core as designated by
   \citet{1993ApJ...406..122A} in their submillimeter continuum maps. We used the
   single-sideband tuned APEX-1 receiver which is part of the Swedish heterodyne
   facility instrument \citep{2008A&A...490.1157V}. It has a sideband rejection
   ratio more than 10~dB. The image band is 12 GHz above or below the observing
   frequency depending on whether the tuning is optimized for operation in the upper or
   lower sidebands, respectively. For two frequency settings, the strong CO(2-1)
   and {\cio}(2-1) lines entered via the image band and from the strength we could
   estimate the sideband rejection ratio to be about 15~dB in  both cases. As
   backend we used two 1 GHz modules of the FFTS \citep{2006A&A...454L..29K}. Each
   FFTS 1 GHz module has 8192 effective channels and the modules can be placed
   independently within the IF band width of 4-8 GHz. At 230 GHz this channel
   spacing corresponds to $0.16\kmps$ and the telescope beamsize (HPBW) is
   $27\arcsec$.
   
   The telescope control software APECS \citep{2006A&A...454L..25M} was used to control
   the raster mapping. All observations were performed in position-switching mode using
   a reference position offset by $300\arcsec$ east and $200\arcsec$ north of the
   map center. The pointing of the telescope was maintained and checked regularly by
   means of small CO(2-1) cross maps of the relatively nearby carbon stars
   IRAS15194$-$5115 and RAFGL1922. The determined pointing offsets were generally consistent from day to day
   and we believe we have an absolute pointing uncertainty better than $5\arcsec$. To
   optimize the antenna focussing we used continuum observations on Jupiter and Saturn.
   
   The observations took place in two blocks and one additional day during 2009:
   April 24 - May 1, May 21, and July 4 - July 9. The column of precipitable water
   vapour was typically around 0.7 mm (varied between 0.3-2.9 mm). Typical system
   temperatures for the frequencies in question (218-252 GHz) were 200-220 K. The
   telescope main beam efficiency is $\etamb = 0.73$ at 345 GHz
   \citep{2006A&A...454L..13G}. Hence, using the antenna surface accuracy of
   $18\micron$ we estimate that (using the Ruze formula) that the APEX main beam
   efficiency around 230 GHz is just slightly higher than at 345 GHz, about 0.75
   which we adopt when converting the observed intensities from the $T_\mathrm{A}^*$
   scale to the $T_\mathrm{mb}$ intensity scale. The heterodyne calibration
   procedure at APEX is more elaborate than the standard chopper wheel calibration
   scheme and involves three measurements by apart from the normal sky observation
   it also measures the receiver temperature by observing a hot and cold loads.
   Moreover, the atmospheric contribution is based on the model
   \citep{2001ITAP...49.1683P} that has been adapted to the atmospheric
   characteristics at the Chajnantor site. The absolute calibration uncertainty is
   estimated to be 10\% in the 1 mm band.

   In \tref{t:hhco_obs} we summarize the targeted formaldehyde lines. The additional
   lines are summarized in \tref{t:other_obs}. In the tables we include the line
   parameters: the transition frequency (typical uncertainty is 0.05 MHz or better),
   energy of lower level, and Einstein $A$-coefficient. These have been compiled
   from the Cologne Database of Molecular Spectroscopy \citep{2001A&A...370L..49M,
   2005JMoSt.742..215M}. We here also indicate the symmetry due to the nuclear spin
   direction of the H (or D) nuclei, which for {\hhco}, {\hhico} and {\ddco} can be
   ortho (parallel spins) or para (anti-parallel spins). The statistical weight
   ratio of the symmetries is also noted if applicable. In the case of {\chhhoh},
   the internal rotation of the methyl group results in the symmetry species A and
   E. Note that radiative or non-reactive collisional transitions are forbidden
   between levels of different symmetry.

\begin{table}
\caption{Observed {\hhco}, {\hhico}, {\hdco}, and {\ddco} lines}
\label{t:hhco_obs}
\centering
\begin{tabular}{lrrrr}
\hline\hline
Frequency & Transition & \mc{$E_l$} & \mc{$A_{ul}$} & Symmetry \\
(MHz)     &            & \mc{(K)}   & \mc{($\mathrm{s}^{-1}$)} \\
\hline
\multicolumn{5}{c}{$\rm H_2CO$ ($o/p = 3$)}\\
218222.19 & $3_{0,3}-2_{0,2}$ & 10.6 & $2.8\times 10^{-4}$ & para \\
218475.63 & $3_{2,2}-2_{2,1}$ & 57.6 & $1.6\times 10^{-4}$ & para \\
218760.07 & $3_{2,1}-2_{2,0}$ & 57.6 & $1.6\times 10^{-4}$ & para \\
225697.78 & $3_{1,2}-2_{1,1}$ & 22.6 & $2.8\times 10^{-4}$ & ortho\\
\multicolumn{5}{c}{$\rm H_2^{13}CO$ ($o/p = 3$)}\\
219908.52 & $3_{1,2}-2_{1,1}$ & 22.4 & $2.6\times 10^{-4}$ & ortho\\
\multicolumn{5}{c}{$\rm HDCO$}\\
227668.45 & $1_{1,1}-0_{0,0}$ &  0.0 & $1.7\times 10^{-6}$ &      \\
246924.60 & $4_{1,4}-3_{1,3}$ & 25.8 & $4.0\times 10^{-4}$ &      \\
\multicolumn{5}{c}{$\rm D_2CO$ ($o/p = 2$)}\\
221191.66 & $4_{1,4}-3_{1,3}$ & 21.3 & $2.9\times 10^{-4}$ & para  \\
231410.23 & $4_{0,4}-3_{0,3}$ & 16.8 & $2.8\times 10^{-4}$ & ortho \\
233650.44 & $4_{2,3}-3_{2,2}$ & 38.4 & $2.7\times 10^{-4}$ & ortho \\
234293.36$^a$ & $4_{3,2}-3_{3,1}$ & 65.4 & $1.6\times 10^{-4}$ & para \\
234331.06$^a$ & $4_{3,1}-3_{3,0}$ & 65.4 & $1.6\times 10^{-4}$ & para \\
245532.75 & $4_{1,3}-3_{1,2}$ & 23.1 & $3.9\times 10^{-4}$ & para  \\
\hline
\multicolumn{5}{l}{$^a$ only observed at $(0,-30\arcsec)$}
\end{tabular}
\end{table}

\begin{table}
\caption{Additional lines}
\label{t:other_obs}
\centering
\begin{tabular}{lrrrr}
\hline\hline
Frequency & Transition & \mc{$E_l$} & \mc{$A_{ul}$} & Symmetry \\
(MHz)     &            & \mc{(K)}   & \mc{($\mathrm{s}^{-1}$)} \\
\hline
\multicolumn{5}{c}{$\rm CH_3OH$ ($A/E = 1$)}\\
218440.05 & $4_{2,2}-3_{1,2}$ & 35.0 & $4.7\times 10^{-5}$ & E \\
241700.22 & $5_{0,5}-4_{0,4}$ & 36.3 & $6.0\times 10^{-5}$ & E \\
241767.22 & $5_{1,5}-4_{1,4}$ & 28.8 & $5.8\times 10^{-5}$ & E \\
241791.43 & $5_{0,5}-4_{0,4}$ & 23.2 & $6.0\times 10^{-5}$ & A \\
241879.07 & $5_{1,4}-4_{1,3}$ & 44.3 & $6.0\times 10^{-5}$ & E \\
241904.15 & $5_{2,4}-4_{2,3}$ & 49.1 & $5.1\times 10^{-5}$ & E \\
241904.65 & $5_{2,3}-4_{2,2}$ & 45.5 & $5.0\times 10^{-5}$ & E \\
\multicolumn{5}{c}{$\rm SO$} \\
219949.44 & $5_6-4_5$ & 24.4 & $1.3\times 10^{-4}$ & \\
251825.77$^a$ & $6_5-5_4$ & 38.6 & $2.0\times 10^{-4}$ & \\
\multicolumn{5}{c}{$\rm ^{34}SO$} \\
246663.47 & $6_5-5_4$ & 38.1 & $1.8\times 10^{-4}$ & \\
\multicolumn{5}{c}{$\rm SO_2$} \\
226300.03 & $14_{3,11}-14_{2,12}$ & 108.1 & $1.1\times 10^{-4}$ \\
241615.80 & $5_{2,4}-4_{1,3}$ & 12.0 & $8.5\times 10^{-5}$ \\
245563.42 & $10_{3,7}-10_{2,8}$ & 60.9 & $1.2\times 10^{-4}$ \\
\multicolumn{5}{c}{$\rm ^{34}SO_2$} \\
219355.01 & $11_{1,11}-10_{0,10}$ & 49.5 & $1.1\times 10^{-4}$ \\
241985.45 & $8_{3,5}-8_{2,6}$ & 42.8 & $1.1\times 10^{-4}$ & \\
246686.12 & $4_{3,1}-4_{2,2}$ & 18.7 & $8.4\times 10^{-5}$ & \\
\multicolumn{5}{c}{$\rm N_2D^+$} \\
231321.83 & $3-2$ & 11.1 & $7.1\times 10^{-4}$ \\
\hline
\multicolumn{5}{l}{$^a$ only observed at $(0,-30\arcsec)$}
\end{tabular}
\end{table}

\section{Results}
\label{s:res}

\begin{figure}
\includegraphics[width=9cm]{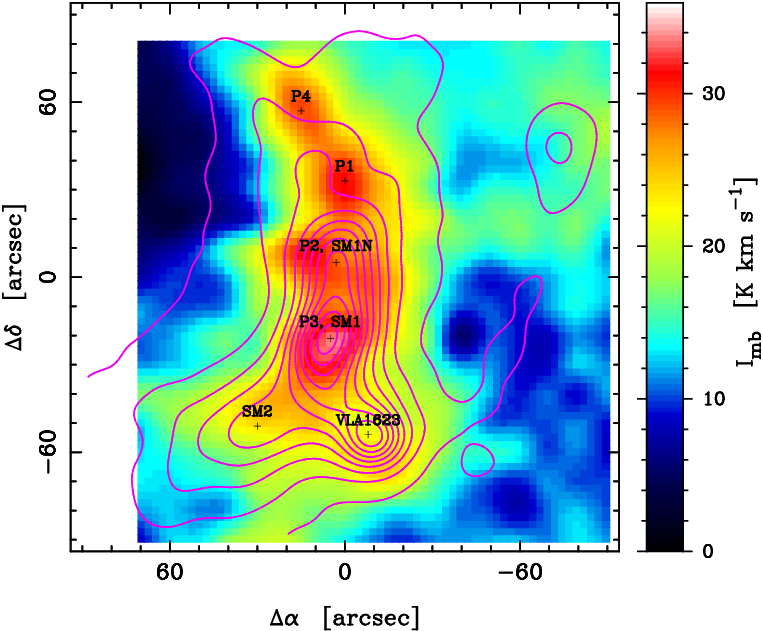}
\caption{Colour image: The {\cio}(3-2) integrated intensity map of
{\roa} cloud core from
\citet{2010A&A...510A..98L}. The intensity scale is shown to the right.
Contours: The 1.3 mm continuum data by \citet{1998A&A...336..150M}. The
contours start at 0.15 Jy with subsequent
contours at every increment of 0.15 Jy. The flux densities are given in a
$15\arcsec$ beam. We show the positions of the mm peaks from
\citet{1998A&A...336..150M} and the {\cio} peaks of \citet{2010A&A...510A..98L}.
The map offsets are given relative the
position \RA{J2000}{16}{26}{27.2} and \Dec{J2000}{-24}{23}{34}.}
\label{f:maps}
\end{figure}

In this section we will first display the formaldehyde mapping results of
the {\roa} cloud core, both as
spectra, integrated intensity maps or velocity position diagrams where
appropriate. After that, the mapping results
for the other molecules will presented. Comparisons will be made with
the existing APEX {\cio}(3-2) data at 329 GHz of \citet{2010A&A...510A..98L} and the
IRAM 30 m continuum map at 1.3 mm of \citet{1998A&A...336..150M}. The angular
resolution of the two data sets is similar; the {\cio}
data have an HPBW of $19\arcsec$ while the 1.3 mm continuum map has an angular
resolution of $15\arcsec$. Both these maps are shown in \fref{f:maps} where the
1.3 mm continuum map is shown in contours on top of a grey-scale image
of the {\cio}(3-2) integrated intensity. The $3'\times 3'$ region shown
in \fref{f:maps} corresponds to the area that has been mapped here in formaldehyde.

\subsection{Formaldehyde results}
\label{s:hhco_res}

All lines listed in \tref{t:hhco_obs} were detected except the \jkk{1}{1,1}{0}{0,0}
transition of {\hdco} at 227 GHz. This transition has a much lower spontaneous rate
coefficient than the other transitions. Two of the {\ddco} lines were not mapped and
were only observed in the $(0\arcsec,-30\arcsec)$ position. The typical rms in a map
spectrum is in the range 0.06-0.1 K.

In \fref{f:h2co_d2co} we plot the {\hhco}($3_{0,3}-2_{0,2}$) and
{\ddco}($4_{0,4}-3_{0,3}$) spectra. They consist of 49 spectra separated by $30\arcsec$
on a 7x7 grid. The {\hhco} lines posess a relatively complicated structure as compared to
the {\ddco} lines. The latter line shows only one velocity component around $3.8\kmps$
with full width at half maximum (FWHM) of $0.7\kmps$ and it is peaking strongly at the
$(0\arcsec,-30\arcsec)$ position which,  within the positional uncertainties, coincides
with the SM1 or P3 position (cf. \fref{f:maps}). We will hereafter refer to this position
as the D-peak. The D-peak is also clearly seen in the integrated intensity
maps of the three different {\hhco} isotopologues (\fref{f:h2co_3maps}).

\begin{figure*}
\centering
\includegraphics{./15012fg02.eps}
\caption{{\hhco}($3_{0,3}-2_{0,2}$) and {\ddco}($4_{0,4}-3_{0,3}$) map
spectra toward the {\roa} cloud. The $(0\arcsec,0\arcsec)$ position is the
same as in \fref{f:maps}.
The vertical scale in each spectrum represents $T_\mathrm{mb}$
scale in K and the horizontal scale is velocity $v_\mathrm{LSR}$ with respect to
local standard of rest in $\kmps$ as shown in the upper rightmost panel.}
\label{f:h2co_d2co}
\end{figure*}

\begin{figure}
\includegraphics[width=8.8cm]{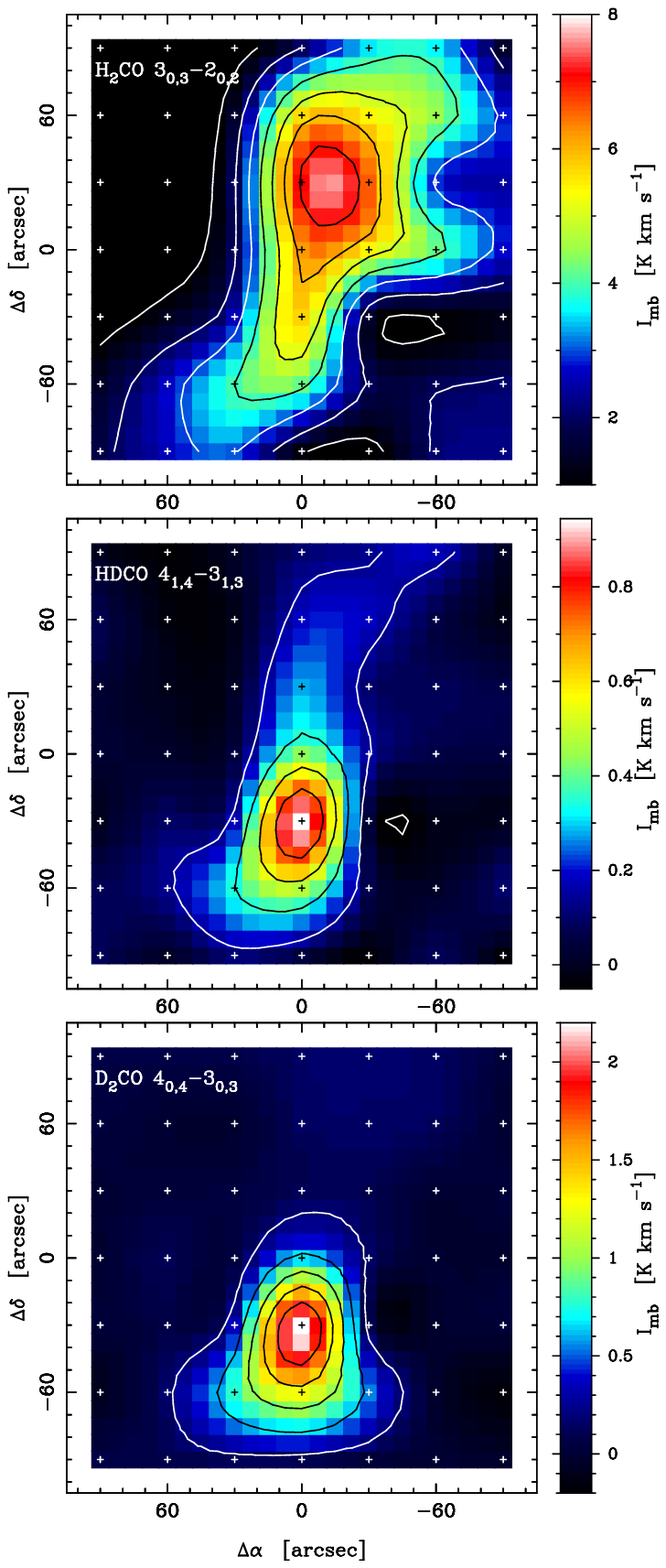}
\caption{Maps of the {\hhco}($3_{0,3}-2_{0,2}$), {\hdco}($4_{1,4}-3_{1,3}$),
and {\ddco}($4_{0,4}-3_{0,3}$) integrated line intensity
over the range $3.0-4.5\kmps$.}
\label{f:h2co_3maps}
\end{figure}

\citet{2004A&A...416..577M} present a {\hhco}($3_{0,3}-2_{0,2}$) IRAM 30~m spectrum
toward the position of VLA1623. The nearest position to VLA1623 in our map is at
$(0\arcsec,-60\arcsec)$. The shape of our spectrum at this position is similar to
the one presented by \citet{2004A&A...416..577M} and our $T_\mathrm{mb}$ peak
temperature of $3.8\K$ is slightly higher than theirs $T_\mathrm{a}^\ast/\etamb
\approx 2.2/0.62 \approx 3.5\K$ despite the higher resolution of $11\arcsec$ in the
IRAM 30 m spectrum. This would be expected for a relatively extended source which
shows little variation on the scale of $11\arcsec - 27\arcsec$. Also, from the
distribution of the {\hhco} emission we do not see any strong component that can be
attributed to VLA1623.\footnote{Based on the radial intensity distributions of the
sub-millimeter continuum, \citet{2001ApJ...548..310J} suggested that the infall zone
around VLA1623 is surrounded by a constant-density region which is the dominating
contribution to the radial intensity profiles at scales $\ga 10\arcsec$.}

The velocity as obtained from the {\nnhp}(1-0) observations by
\citet{2004ApJ...617..425D, 2009ApJ...700.1994D} \citep[see
also][]{2007A&A...472..519A} toward SM1 is $3.7\kmps$ with an FWHM of $0.6\kmps$,
i.e. very similar to our values determined from the {\ddco} lines at the D-peak. The
peak intensity of the {\ddco} line is about half that of {\hhco}. The size of
{\ddco} $4_{0,4}-3_{0,3}$ emission region (see \fref{f:h2co_3maps}), is found, by
fitting a 2-dimensional gaussian, to be $33\arcsec\times 61\arcsec$.  This size
corresponds to a deconvolved source size of about $S= 19\arcsec\times 55\arcsec$ for a
beam size of $B=27\arcsec$. The filling factor when pointing the $27\arcsec$ beam toward
the center of the source, is then $S^2/(S^2+B^2) =
(19\arcsec\times 55\arcsec)/(33\arcsec\times 61\arcsec)\approx 0.5$.
This filling factor is merely an upper limit because any unresolved small-scale
clumpiness may decrease the filling factor further.

In order to investigate the {\hhco} line distribution over the {\roa} cloud core we
plot four velocity position diagrams (\fref{f:vp_diagrams}) between $\Delta\delta
= \pm 90\arcsec$ at $\Delta\alpha =0\arcsec$. Here we also see that HDCO is peaking
at the D-peak at $\Delta\delta = -30\arcsec$. Looking at the {\hdco} emission,
there appears to be a velocity gradient over the D-peak source of about $0.2\kmps$
when going from $\Delta\delta = 0\arcsec$ to $\Delta\delta = -60\arcsec$. It is
also close to the D-peak where the {\cio}(3-2) emission has its maximum. The
secondary {\cio}(3-2) peak (P1 in \fref{f:maps}) is at $\Delta\delta = +30\arcsec$
and with a lower velocity of $\approx 3.1\kmps$ as compared to the velocity of
$+3.7\kmps$ at the D-peak position and it is also here where the low-energy {\hhco}
line intensities reach their maximum value at $\vlsr = 3.3\kmps$. We denote this
peak P1 from now on. There is also weak emission emanating from both {\hdco} and
{\ddco} at the $\Delta\delta = +30\arcsec$ position.

\onlfig{4}{
\begin{figure*}
\centering
\includegraphics{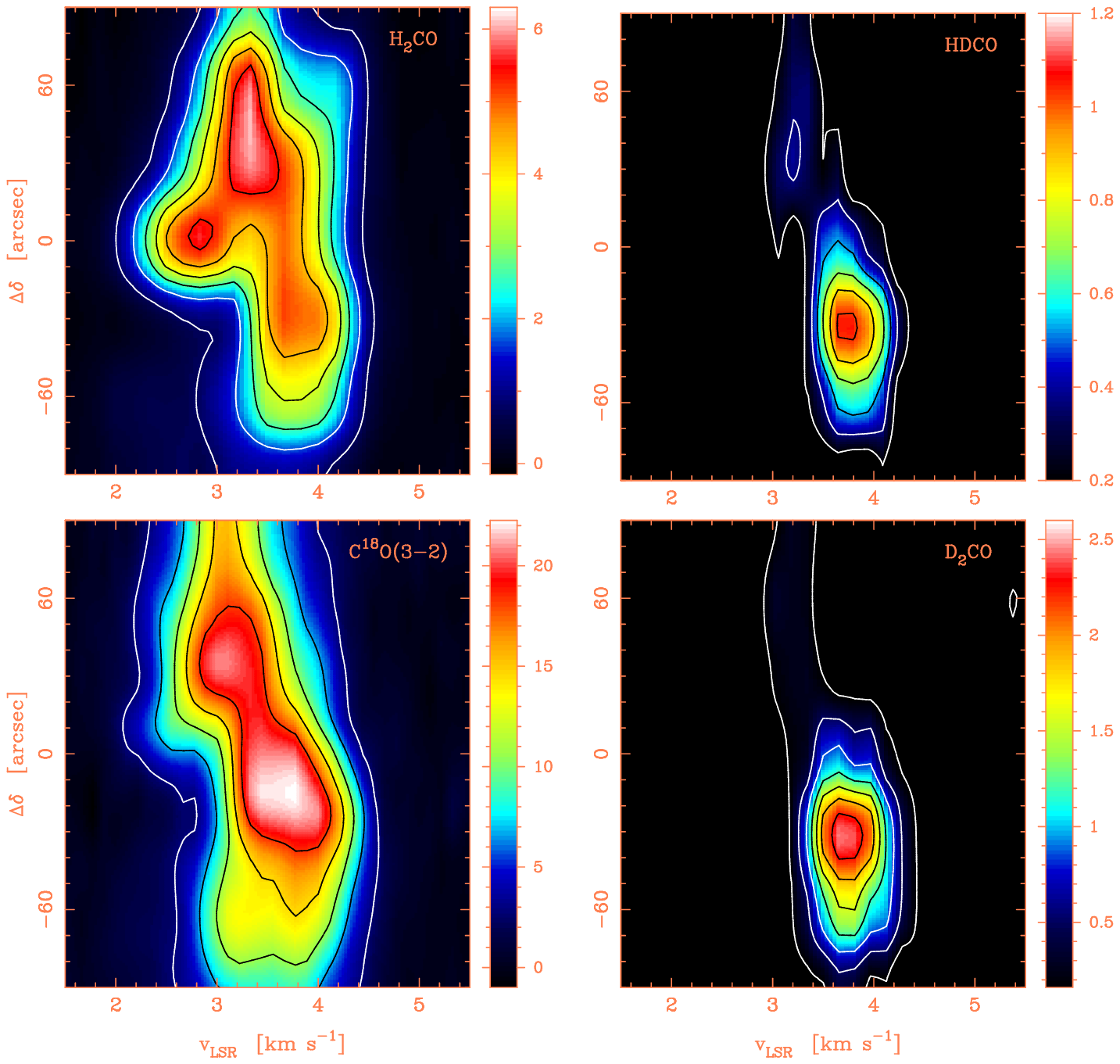}
\caption{Velocity position diagrams of the {\hhco}($3_{0,3}-2_{0,2}$),
{\hdco}($4_{1,4}-3_{1,3}$), {\ddco}($4_{0,4}-3_{0,3}$), and {\cio}(3-2) lines
along the declination axis at $\Delta\alpha=0''$. The molecule is indicated in each
panel. The {\cio}(3-2) data are from \citet{2010A&A...510A..98L}. The vertical
scale in each panel indicates the declination offset and the horizontal scale
is the $v_\mathrm{LSR}$ velocity. The intensity scale in all four panels is
$T_\mathrm{mb}$.}
\label{f:vp_diagrams}
\end{figure*}
}

Interestingly, there is a third {\hhco} peak at $\Delta\delta = 0\arcsec$ and
$v_\mathrm{LSR}\approx +2.9\kmps$ which has no obvious counterpart in the {\cio}
velocity position map. This peak makes the {\hhco}($3_{0,3}-2_{0,2}$) spectrum at
$(0\arcsec,0\arcsec)$ to look doubly peaked (\fref{f:h2co_d2co}). However, these are
two different cloud components and the dip is not an effect of self-absorption. This
is evident from the velocity position diagrams in \fref{f:vp_diagrams}, but is also
clear when comparing the {\hhico}($3_{1,2}-2_{1,1}$) and {\hhco}($3_{1,2}-2_{1,1}$)
spectra. In \fref{f:12_13} the central three spectra from the ortho lines
{\hhco}($3_{1,2}-2_{1,1}$) and {\hhico}($3_{1,2}-2_{1,1}$) are displayed. Toward the
$(0\arcsec,0\arcsec)$ position we see the doubly peaked line profile also for this
{\hhco} line, however there is no hint that the {\hhico} line is peaking at the
velocity of the dip for {\hhco}. If anything, the emission from the rarer species
seems to follow that of the main species.
We will return
to the low-velocity component seen in the $(0\arcsec,0\arcsec)$ position below when
we present the {\chhhoh} results. 

\onlfig{5}{
\begin{figure*}
\centering
\includegraphics[width=12.0cm]{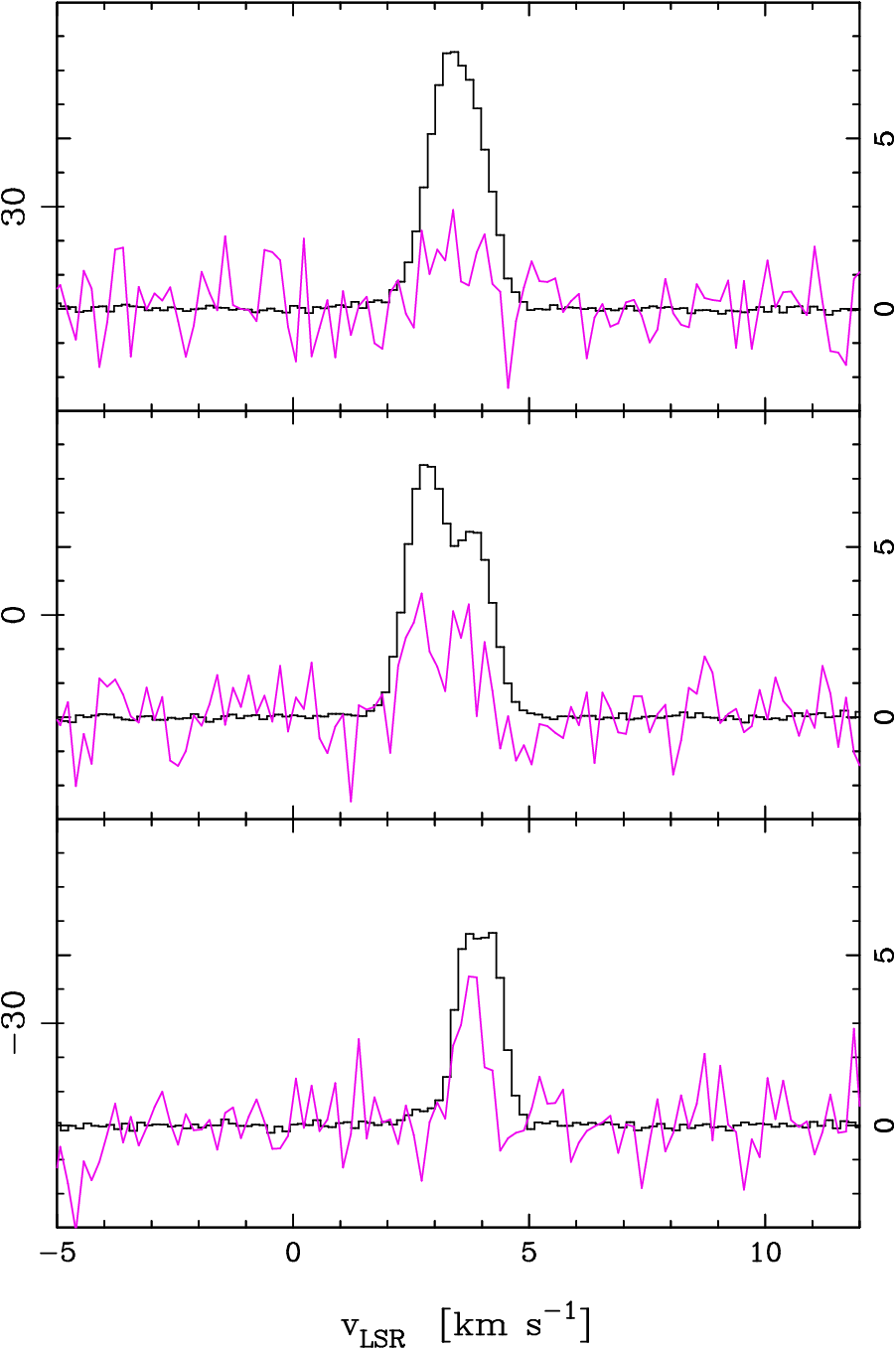}
\caption{Ortho {\hhco}($3_{1,2}-2_{1,1}$) and {\hhico}($3_{1,2}-2_{1,1}$) spectra at
declination offsets $-30\arcsec$ (bottom, D-peak), $0\arcsec$ (middle) and $+30\arcsec$
(top, P1). The offsets are relative the center position in \fref{f:maps}. The {\hhico}
spectra (in colour) have been scaled up with a factor of 10 and appear much more
noisy.}
\label{f:12_13}
\end{figure*}
}

\subsection{{\chhhoh} results}
\label{s:ch3oh_res}
\begin{figure}
\includegraphics[width=8.8cm]{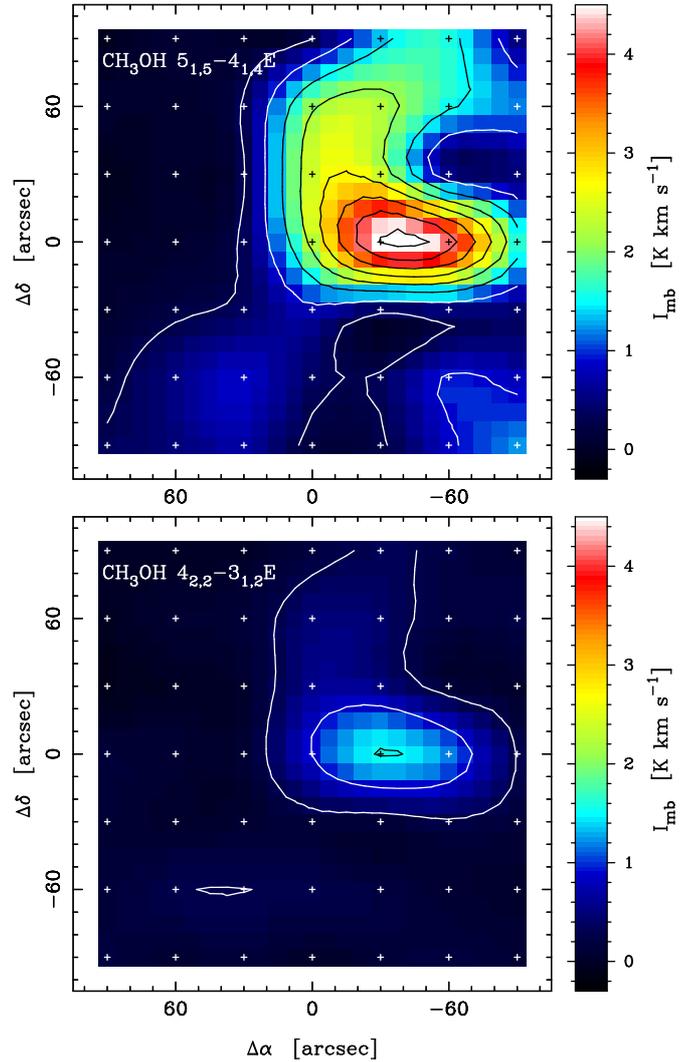}
\caption{Integrated intensity maps of the {\chhhoh} $5_{1,5}-4_{1,4}$ and
$4_{2,2}-3_{1,2}$ E-lines. First contour
is at $0.3\Kkmps$ and the increment is $0.6\Kkmps$.}
\label{f:ch3oh_maps}
\end{figure}

\begin{figure}
\includegraphics[width=8.8cm]{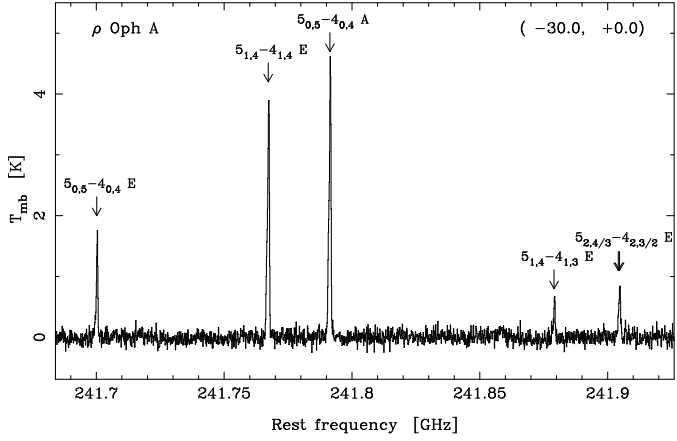}
\caption{{\chhhoh} 5-4 lines around 241 GHz toward the $(-30\arcsec,0\arcsec)$
position. Note that the $5_{2,4}-4_{2,3}$ and $5_{2,3}-4_{2,2}$ E-lines are
blended.}
\label{f:ch3oh_241}
\end{figure}

In \fref{f:ch3oh_maps} we show a map of the integrated intensity for the
{\chhhoh} $5_{1,5}-4_{1,4}$ and $4_{2,2}-3_{1,2}$ E-lines. The latter line
comes from levels of higher energy (\tref{t:hhco_obs}). The distribution of {\chhhoh} is quite
different from that of {\hhco} (see \fref{f:h2co_3maps}). It should be noted that
the {\chhhoh} $4_{2,2}-3_{1,2}$ line and the {\hhco} lines at 218 GHz were
observed simultaneously so the different distributions cannot be a result
of large pointing offsets. The {\chhhoh}
emission has its maximum at $(-30\arcsec,0)$ and in \fref{f:ch3oh_241}
all observed $5-4$ lines in this position are shown. Here the peak emission
velocity is lower, around $\vlsr\approx 2.8\kmps$. The cloud component
at $(-30\arcsec,0\arcsec)$ extends into adjacent positions. The low-velocity feature
seen for {\hhco} (\fref{f:vp_diagrams}) is very likely
associated with the {\chhhoh}-peak. The higher energy {\hhco} lines $3_{2,2/1}-2_{2,1/0}$
peak near this position. In addition, there is little {\chhhoh} at the
D-peak position (\fref{f:ch3oh_maps}).

\subsection{Results from other molecules}
\label{s:res_oth}
Two lines from {\so}($5_6-4_5$) and {\iso}($6_5-5_4$) were covered during the
observations. The map spectra are shown in \fref{f:so_34so}. There is a single
strong peak in the {\iso} line intensity toward the position $(-60\arcsec,
+60\arcsec)$ which we will call the S-peak. Here the {\iso} and {\soo} lines are
very narrow, only about $0.4\kmps$. This is about twice the width one would expect
from thermal broadening only (at $\tkin=20\K$). Further west, at $(-90\arcsec,
+30\arcsec)$, the {\iso} line is broader, about $1.1\kmps$. The profile of {\so}
shows line wings here which could be related to outflow activity. However, it is at
the very border of our map and a clear delineation into red and blue wings are
impossible. It should be pointed out that the S-peak has no structural counterpart
in the {\cio}(3-2) map nor the 1.3 mm continuum map (cf. \fref{f:maps}). Toward the
D-peak, the {\so} lines exhibit a similar emission velocity ($\vlsr\approx
3.7\kmps$) and line width ($\dv\approx 0.83\kmps$) as the other molecules. The
covered {\soo} and {\isoo} lines (\tref{t:other_obs}) show a distribution
similar to that of {\so} and {\iso}. At the P1-position the {\so} and
{\soo} emission velocity of about $\approx 2.9\kmps$ is lower than the velocity of
$\approx 3.3\kmps$ for the formaldehyde isotopologues and {\chhhoh}.

\onlfig{9}{
\begin{figure*}
\centering
\includegraphics{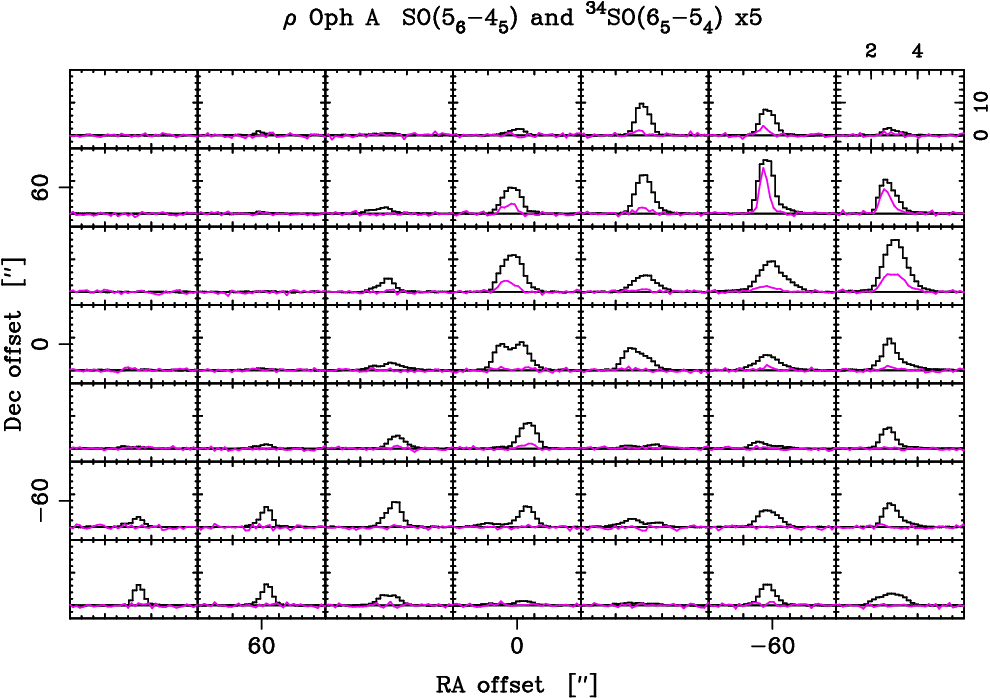}
\caption{{\so}($5_6-4_5$) and {\iso}($6_5-5_4$) map
spectra toward the {\roa} cloud. The {\iso} spectra (in colour) have been multiplied with
a factor of 5. Scales as in \fref{f:h2co_d2co}.}
\label{f:so_34so}
\end{figure*}
}

\begin{figure}
\includegraphics[width=8.8cm]{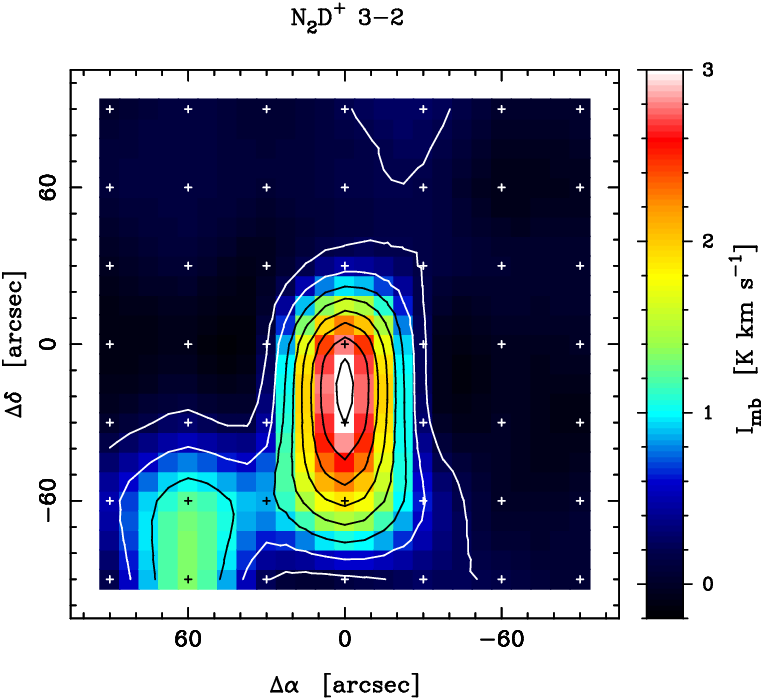}
\caption{Integrated intensity map of the {\nndp} 3-2 line toward {\roa}. First
contour is at $0.2\Kkmps$ and the increment is $0.4\Kkmps$.}
\label{f:nndp_map}
\end{figure}

The deuterated version of {\nnhp} \cite[observed by][]{2004ApJ...617..425D,
2009ApJ...700.1994D, 2007A&A...472..519A}, {\nndp}(3-2) at 231 GHz was also covered
during the formaldehyde observations. The integrated intensity map of {\nndp}(3-2)
is shown in \fref{f:nndp_map}. The peak intensity of the {\nndp} emission is clearly
associated with the D-peak, with a secondary weaker source to the south-east. This
secondary peak is not coincident with SM2 (see \fref{f:maps}) but is located in the
direction where the dust emission extends. Moreover, it shows up in the {\nnhp} map
of \citet{2004ApJ...617..425D} and is designated N6 by them. The emission velocity
of {\nndp}(3-2) at the D-peak is $3.7\kmps$ and the FWHM is $0.85\kmps$ in agreement
with what is found for the other species.

\subsection{Result summary}
\label{s:res_sum}
After presenting the results of the distribution of the different molecules toward
the {\roa} cloud above we will now make a summary of the results which will be used
in the analysis below. For {\hhco} and {\chhhoh} we have identified three positions
of interest which seem to represent distinctive sources; the D-peak at
$(0\arcsec,-30\arcsec)$, the P1 peak at $(0\arcsec,+30\arcsec)$, and the {\chhhoh}
peak at $(-30\arcsec, 0\arcsec)$. In addition to these three peaks, the sulphur
containing molecules are localized to $(-60\arcsec, +60\arcsec)$ (the S-peak). The
emission velocities and FWHMs as obtained for the different species in these sources
are summarized in \tref{t:peaks} (from gaussian fits of the lowest energy lines
in those cases where we have mapped multiple lines). Integrated line intensities
for the relevant positions are tabulated in \tref{t:hhco_intint},
\tref{t:ch3oh_intint}, and \tref{t:other_intint}. In these tables the $1\sigma$
uncertainties of the integrated intensities due to noise have been entered. 

\begin{table}
\caption{Properties of the {\hhco}, {\chhhoh}, {\so}, and {\soo} emission
in selected cloud positions from gaussian fits}
\label{t:peaks}
\centering
\begin{tabular}{lrrr}
\hline\hline
Offset & Molecule & $v_\mathrm{LSR}$ & $\Delta v$ (FWHM) \\
       &          & ($\kmps$)        & ($\kmps$) \\
\hline
$(0\arcsec,-30\arcsec)$    & {\hhco}   & 3.8(0.1) & 0.95(0.05) \\
                 & {\hdco}   & 3.6(0.1) & 0.70(0.05) \\
		 & {\ddco}   & 3.8(0.1) & 0.68(0.02) \\
		 & {\chhhoh} & 3.7(0.1) & 0.73(0.05) \\
                 & {\so}     & 3.7(0.1) & 0.83(0.05) \\
                 & {\soo}    & 3.7(0.1) & 0.67(0.05) \\
$(0\arcsec,+30\arcsec)$    & {\hhco}   & 3.4(0.1) & 1.20(0.05) \\
                 & {\hdco}   & 3.2(0.1) & 0.85(0.12) \\
		 & {\ddco}   & 3.2(0.1) & 0.60(0.15) \\
		 & {\chhhoh} & 3.4(0.1) & 1.18(0.05) \\
                 & {\so}     & 3.0(0.1) & 1.06(0.05) \\
                 & {\soo}    & 2.8(0.1) & 0.77(0.05) \\
$(-30\arcsec,0\arcsec)^a$    & {\hhco}   & 2.8(0.2) & 0.95(0.10) \\
                 &           & 3.6(0.2) & 0.87(0.10) \\
		 & {\chhhoh} & 2.6(0.2) & 0.62(0.10) \\
		 &           & 3.2(0.2) & 1.27(0.10) \\
$(-60\arcsec, +60\arcsec)$ & {\so} &  3.0(0.1) &  0.78 (0.05) \\
                 & {\iso}    &  2.9(0.1) & 0.43(0.02) \\
                 & {\soo}    &  2.9(0.1) & 0.42(0.02) \\
                 & {\isoo}   &  3.0(0.1) & 0.35(0.02) \\
\hline
\multicolumn{4}{l}{$^a$ Fitted by two components, full intensity used in models}
\end{tabular}
\end{table}

\begin{table}
\caption{{\hhco}, {\hhico}, {\hdco}, and {\ddco}
integrated line intensities in three positions of the {\roa} cloud}
\label{t:hhco_intint}
\centering
\begin{tabular}{lrrrr}
\hline\hline
Molecule & Line & Frequency &  $I_\mathrm{mb}=\int T_\mathrm{mb}\, dv$ & $\Delta I_\mathrm{mb}$ $^a$\\
         &      & (MHz)    & ($\Kkmps$) & ($\Kkmps$)       \\
\hline
\multicolumn{5}{c}{D-peak $(0\arcsec,-30\arcsec)$} \\
{\hhco}   & $3_{0,3}-2_{0,2}$ & 218222 &  6.07 & 0.05 \\
          & $3_{2,2}-2_{2,1}$ & 218476 &  0.83 & 0.05 \\
          & $3_{2,1}-2_{2,0}$ & 218760 &  0.80 & 0.05 \\
          & $3_{1,2}-2_{1,1}$ & 225698 &  6.88 & 0.06 \\
{\hhico}  & $3_{1,2}-2_{1,1}$ & 219908 &  0.26 & 0.06 \\
{\hdco}   & $4_{1,4}-3_{1,3}$ & 246924 &  0.93 & 0.08 \\
{\ddco}   & $4_{1,4}-3_{1,3}$ & 221192 &  0.87 & 0.08 \\ 
          & $4_{0,4}-3_{0,3}$ & 231410 &  2.21 & 0.05 \\
          & $4_{2,3}-3_{2,2}$ & 233650 &  0.45 & 0.005 \\
          & $4_{3,2}-3_{3,1}$ & 234293 &  0.032 & 0.005 \\
          & $4_{3,1}-3_{3,0}$ & 234331 &  0.028 & 0.005 \\
          & $4_{1,3}-3_{1,2}$ & 245533 &  0.57  & 0.05 \\
\multicolumn{5}{c}{P1 $(0\arcsec,+30\arcsec)$} \\
{\hhco}   & $3_{0,3}-2_{0,2}$ & 218222 &  8.69 & 0.05 \\
          & $3_{2,2}-2_{2,1}$ & 218476 &  1.28 & 0.06 \\
          & $3_{2,1}-2_{2,0}$ & 218760 &  1.19 & 0.06 \\
          & $3_{1,2}-2_{1,1}$ & 225698 & 10.96 & 0.07 \\
{\hhico}  & $3_{1,2}-2_{1,1}$ & 219908 &  0.26 & 0.07 \\
{\hdco}   & $1_{1,1}-0_{0,0}$ & 227668 &       & 0.06 \\
          & $4_{1,4}-3_{1,3}$ & 246924 &  0.44 & 0.05 \\
{\ddco}   & $4_{1,4}-3_{1,3}$ & 221192 &  0.13 & 0.05 \\ 
          & $4_{0,4}-3_{0,3}$ & 231410 &  0.19 & 0.05 \\
\multicolumn{5}{c}{{\chhhoh}-peak $(-30\arcsec, 0\arcsec)$} \\
{\hhco}   & $3_{0,3}-2_{0,2}$ & 218222 &  8.56 & 0.05 \\
          & $3_{2,2}-2_{2,1}$ & 218476 &  1.45 & 0.06 \\
          & $3_{2,1}-2_{2,0}$ & 218760 &  1.41 & 0.06 \\
          & $3_{1,2}-2_{1,1}$ & 225698 & 10.13 & 0.07 \\
{\hhico}  & $3_{1,2}-2_{1,1}$ & 219908 &  0.27 & 0.06 \\
{\hdco}   & $4_{1,4}-3_{1,3}$ & 246924 &  0.12 & 0.05 \\
{\ddco}   & $4_{1,4}-3_{1,3}$ & 221192 &  0.12 & 0.05 \\ 
          & $4_{0,4}-3_{0,3}$ & 231410 &  0.15 & 0.05 \\
\hline
\multicolumn{5}{l}{$^a$ 1$\sigma$ error due to noise only}
\end{tabular}
\end{table}

\begin{table}
\caption{{\chhhoh}
integrated line intensities in three positions of the {\roa} cloud}
\label{t:ch3oh_intint}
\centering
\begin{tabular}{lrrr}
\hline\hline
Line & Frequency &  $I_\mathrm{mb}=\int T_\mathrm{mb}\, dv$ & $\Delta I_\mathrm{mb}$ $^a$\\
     & (MHz)	& ($\Kkmps$) & ($\Kkmps$)	\\
\hline
\multicolumn{4}{c}{D-peak $(0\arcsec,-30\arcsec)$} \\
$4_{2,2}-3_{1,2}$E1 & 218440 & 0.21 & 0.05 \\
$5_{0,5}-4_{0,4}$E1 & 241700 & 0.13 & 0.05 \\
$5_{1,5}-4_{1,4}$E2 & 241767 & 0.72 & 0.05 \\
$5_{0,5}-4_{0,4}$A  & 241791 & 0.81 & 0.05 \\
$5_{1,4}-4_{1,3}$E1 & 241879 & 0.09 & 0.05 \\
$5_{2,3/4}-4_{2,2/3}$E1 & 241904 & 0.07 & 0.05 \\
\multicolumn{4}{c}{P1 $(0\arcsec,+30\arcsec)$} \\
$4_{2,2}-3_{1,2}$E1 & 218440 & 0.61 & 0.05 \\
$5_{0,5}-4_{0,4}$E1 & 241700 & 0.59 & 0.05 \\
$5_{1,5}-4_{1,4}$E2 & 241767 & 2.39 & 0.05 \\
$5_{0,5}-4_{0,4}$A  & 241791 & 2.88 & 0.05 \\
$5_{1,4}-4_{1,3}$E1 & 241879 & 0.20 & 0.05 \\
$5_{2,3/4}-4_{2,2/3}$E1 & 241904 & 0.31 & 0.05 \\
\multicolumn{4}{c}{{\chhhoh}-peak $(-30\arcsec, 0\arcsec)$} \\
$4_{2,2}-3_{1,2}$E1 & 218440 & 1.43 & 0.05 \\
$5_{0,5}-4_{0,4}$E1 & 241700 & 1.49 & 0.05 \\
$5_{1,5}-4_{1,4}$E2 & 241767 & 4.53 & 0.05 \\
$5_{0,5}-4_{0,4}$A  & 241791 & 5.25 & 0.05 \\
$5_{1,4}-4_{1,3}$E1 & 241879 & 0.96 & 0.05 \\
$5_{2,3/4}-4_{2,2/3}$E1 & 241904 & 1.43 & 0.05 \\
\hline
\multicolumn{4}{l}{$^a$ 1$\sigma$ error due to noise only}
\end{tabular}
\end{table}

\begin{table}
\caption{Integrated line intensities for other molecules in four positions
of the {\roa} cloud}
\label{t:other_intint}
\centering
\begin{tabular}{lrrr}
\hline\hline
Transition & Frequency &  $I_\mathrm{mb}=\int T_\mathrm{mb}\, dv$ & $\Delta I_\mathrm{mb}$ $^a$\\
            & (MHz)     & ($\Kkmps$) & ($\Kkmps$) \\
\hline
\multicolumn{4}{c}{D-peak $(0\arcsec,-30\arcsec)$} \\
\so($5_6-4_5$)              & 219949 & 7.22 & 0.06 \\
\so($6_5-5_4$)              & 251826 & 3.37 & 0.04 \\
\iso($6_5-5_4$)             & 246663 & 0.19 & 0.03 \\
\soo($5_{2,4}-4_{1,3}$)     & 241616 & 0.69 & 0.02 \\
\soo($10_{3,7}-10_{2,8}$)   & 245563 & 0.16 & 0.04 \\
\nndp($3-2$)                & 231322 & 3.38 & 0.06 \\
\multicolumn{4}{c}{P1 $(0\arcsec,+30\arcsec)$} \\
\so($5_6-4_5$)              & 219949 & 13.1 & 0.06 \\
\iso($6_5-5_4$)             & 246663 & 0.55 & 0.04 \\
\soo($5_{2,4}-4_{1,3}$)     & 241616 & 1.70 & 0.05 \\
\soo($10_{3,7}-10_{2,8}$)   & 245563 & 0.57 & 0.04 \\
\nndp($3-2$)                & 231322 & 0.37 & 0.08 \\
\multicolumn{4}{c}{{\chhhoh}-peak $(-30\arcsec, 0\arcsec)$} \\
\so($5_6-4_5$)              & 219949 & 8.25 & 0.06 \\
\iso($6_5-5_4$)             & 246663 & 0.08 & 0.04 \\
\soo($5_{2,4}-4_{1,3}$)     & 241616 & 0.23 & 0.05 \\
\soo($10_{3,7}-10_{2,8}$)   & 245563 & 0.05 & 0.04 \\
\nndp($3-2$)                & 231322 & 0.17 & 0.07 \\
\multicolumn{4}{c}{S-peak $(-60\arcsec, +60\arcsec)$} \\
\so($5_6-4_5$)              & 219949 & 14.5 & 0.06 \\
\iso($6_5-5_4$)             & 246663 & 1.31 & 0.04 \\
\soo($5_{2,4}-4_{1,3}$)     & 241616 & 3.76 & 0.05 \\
\soo($10_{3,7}-10_{2,8}$)   & 245563 & 1.14 & 0.04 \\
\soo($14_{3,11}-14_{2,12}$) & 226300 & 0.27 & 0.03 \\
\isoo($4_{3,1}-4_{2,2}$)    & 246686 & 0.24 & 0.04 \\
\isoo($8_{3,5}-8_{2,6}$)    & 241985 & 0.18 & 0.04 \\
\isoo($11_{1,11}-10_{0,10}$) & 219355 & 0.46 & 0.04 \\
\nndp($3-2$)                & 231322 &       & 0.07 \\
\hline
\multicolumn{4}{l}{$^a$ 1$\sigma$ error due to noise only}
\end{tabular}
\end{table}

\section{Analysis}
\label{s:ana}
For several of the detected species (\tref{t:hhco_intint}) we have multiple
transitions that have significantly different energy levels. In such cases the
so-called rotation diagram analysis \citep[eg.][]{1999ApJ...517..209G} can be
employed to determine the rotation temperature, ${\trot}$, and the molecular
column density, $\nmol$. If all lines are optically thin, the rotation diagram
method often is adequate to analyse multi-transition data. However, here it is
likely that we have a mixture of optically thin and thick lines and therefore we
adopt the modified approach to the rotation diagram method described by
\citet{2000ApJS..128..213N}. The modification involves the inclusion of the peak
optical depth and thus, in addition to $\trot$ and $\nmol$, also the beam filling
factor, $\ebf$, can be determined by minimization of a $\chi^2$-value. For a
gaussian source distribution, $\ebf$ is given by $\ebf = S^2 / (S^2 + B^2)$, where
$S$ is the FWHM source size and $B$ is the FWHM beam size. If all transitions
included in the analysis are optically thin and $\ebf$ is set to 1 ($B << S$), the
method is very similar\footnote{The only difference is that in the normal rotation
diagram analysis a straight line is least-square fitted to quantities proportional
to the logarithm of the line intensities but here the fit is performed directly by
minimizing the sum of the squared and error-weighted differences of observed and
modelled line intensities.} to the rotation diagram analysis with $\nmol$ then
representing a beam averaged column density.

In selected cases we will also check the results and refine the models obtained
with the modified rotation diagram analysis by employing a more accurate treatment
of the line excitation and radiative transfer. We have here adopted the
accelerated lambda iteration (ALI) technique outlined by
\citet{1991A&A...245..171R, 1992A&A...262..209R}. The ALI model cloud consists of
spherically concentric shells and allows only for radial gradients of the physical
parameters (kinetic temperature, molecular hydrogen density, molecular abundance,
and radial velocity field). Moreover, and in contrast to the rotation diagram
method, collisional excitation is included in the analysis, so the collision
partner (here {\hh}) density is a physical input parameter. The code used in the
present work has been tested by \citet{2008A&A...479..779M} and it allows the
inclusion of dust as a source of continuum emission in the shells. When the
radiative transfer has been iteratively solved using the ALI approach, a model
spectrum is produced by convolving the velocity dependent intensity distribution
of the model cloud with a gaussian beam.

\subsection{Analysis of the formaldehyde lines}
\label{s:h2co_ana}
As can be seen in \tref{t:hhco_intint} all four {\hhco} lines are clearly detected
in the three sources (D-peak, P1, and {\chhhoh}-peak). The three para lines at 218
GHz have all been observed simultaneously so their relative strengths are not
affected by any pointing or calibration uncertainties. The only ortho transition,
the $3_{1,2}-2_{1,1}$ line at 225 GHz, was observed in a separate frequency setting.
In our rotation diagram analysis we treat the ortho and para {\hhco} lines together
by assuming that the population distribution is determined by $\trot$ also between
the states of different symmetry (through some exchange reaction or formation
mechanism). For {\hhco} the lowest ortho level ($1_{1,1}$) is about 15 K above the
lowest para level ($0_{0,0}$). For {\ddco} the reverse situation is in effect and it
is the lowest level $0_{0,0}$ that is an ortho state while the lowest para level
$1_{1,1}$ is 8 K higher up in energy. For a very low $\trot$ (during molecule
formation) most of the molecules will be in the lowest energy state (para for
{\hhco} and ortho for {\ddco}). On the other hand, if $\trot$ is much greater than
the energy difference of the symmetry states, the ortho to para population ratio
will be governed by the statistical weight ratio which is $o/p=3$ for {\hhco} and
$o/p=2$ for {\ddco}.

In \tref{t:h2co_rotdia} the results of the modified rotation diagram analysis are
shown. Here the number of lines used in the analysis for each molecule is
tabulated together with best fit values of $\trot$, $\nmol$, and $\ebf$. In the
last column, we list the minimum and maximum optical depths of the used
transitions in the analysis. If there are not enough of lines or all lines are
optically thin, one or two of the parameters have been set to a result obtained by
another molecule in a previous fit. For instance, in the D-peak source, all three
parameters could be determined in the analysis of {\hhco}, while for the optically
thin {\ddco} lines, $\ebf$ had to be set to the value found for {\hhco}. Using the
same $\ebf$ for all formaldehyde isotopologues will also make the determined
column densities directly comparable with each other. Interestingly, the $\trot$
obtained for {\ddco}, $17.4\K$ toward the D-peak is lower than that found for
{\hhco} ($\trot=22.5\K$). We assume this is an effect of subthermal excitation and
difference in optical depths, where the higher optical depths for {\hhco} make the
excitation more efficient via photon trapping as compared to {\ddco}.
A lower rotation temperature of the less abundant formaldehyde isotopologues
was also seen in IRAS16293$-$2422 by \citet{2000A&A...359.1169L} and in several
other sources by \citet{2006A&A...453..949P}.
For the other two sources (the P1 and {\chhhoh} peaks) the number of detected {\ddco}
transitions is not sufficient to allow for a $\trot$ determination and we assume that
the ratio of $\trot$(\ddco)/$\trot$(\hhco) is the same in these two sources as in the
D-peak source. The deduced $\trot$ is higher in the P1 and {\chhhoh} peaks than in the
D-peak. This could be expected because the ratio of the $3_{0,3}-2_{2,0}$ and
$3_{2,2}-2_{2,1}$ {\hhco} lines is a good measure of kinetic temperature
\citep{1993ApJS...89..123M} and this ratio is highest toward the {\chhhoh}-peak. Of
course, from the models we can also estimate the intensity of lines not included in
the analysis. In particular, we find for the {\hdco} $1_{1,1}-0_{0,0}$ line that for
the determined model parameters toward the D-peak its expected intensity is about 5
times below the noise level, consistent with our non-detection. 

\begin{table}
\caption{Analysis results for {\hhco}, {\hhico}, {\hdco}, {\ddco}, {\chhhoh}, {\so},
and {\soo} from using the modified rotation diagram technique}
\label{t:h2co_rotdia}
\centering
\begin{tabular}{lrrrrr}
\hline\hline
Molecule & No. of & $\trot$  & $\nmol$  & $\ebf$ & $\tau_\mathrm{min},\tau_\mathrm{max}$ \\
         & lines  & (K)     & $\mathrm{cm^{-2}}$ \\
\hline
\multicolumn{6}{c}{D-peak $(0\arcsec,-30\arcsec)$} \\
{\hhco}    &  4         &  22.5   & $\me{2.22}{14}$ & 0.446   & 0.11,2.70 \\
{\ddco}    &  6         &  17.4   & $\me{3.17}{13}$ & (0.446) & 0.01,0.54 \\
{\hdco}    &  1         &  (17.4) & $\me{2.37}{13}$ & (0.446) & 0.23 \\
{\hhico}   &  1         &  (17.4) & $\me{4.09}{12}$ & (0.446) & 0.06 \\
{\chhhoh}  &  6         &  6.8    & $\me{7.21}{14}$ & 1       & 0.02,0.62 \\
{\so}      &  2         &  18.3   & $\me{2.38}{14}$ & (1)     & 0.50,1.21 \\
{\iso}     &  1         &  (18.3) & $\me{9.67}{12}$ & (1)     & 0.02 \\
{\soo}     &  2         &  19.9   & $\me{4.23}{13}$ & (1)     & 0.01,0.07 \\
\multicolumn{6}{c}{P1 $(0\arcsec,+30\arcsec)$} \\
{\hhco}    &  4         &  25.9   & $\me{1.90}{14}$ & 0.469   & 0.10,1.91 \\
{\ddco}    &  2         &  (20)   & $\me{1.66}{12}$ & (0.469) & 0.01,0.02 \\
{\hdco}    &  1         &  (20)   & $\me{6.21}{12}$ & (0.469) & 0.05 \\
{\hhico}   &  1         &  (20)   & $\me{1.71}{12}$ & (0.469) & 0.02 \\
{\chhhoh}  &  6         &   7.4   & $\me{1.24}{15}$ & 1       & 0.05,1.20 \\
{\soo}     &  2         &  23.4   & $\me{9.91}{13}$ & (1)     & 0.04,0.12 \\
\multicolumn{6}{c}{{\chhhoh}-peak $(-30\arcsec, 0\arcsec)$} \\
{\hhco}    &  4         &  26.3   & $\me{2.66}{14}$ & 0.325   & 0.14,2.60 \\
{\ddco}    &  2         &  (20)   & $\me{1.93}{12}$ & (0.325) & 0.01,0.03 \\
{\hdco}    &  1         &  (20)   & $\me{2.38}{12}$ & (0.325) & 0.02 \\
{\hhico}   &  1         &  (20)   & $\me{3.58}{12}$ & (0.325) & 0.05 \\
{\chhhoh}  &  6         &   9.2   & $\me{1.26}{15}$ & 1       & 0.11,1.46 \\
{\soo}     &  1         &  (20)   & $\me{8.33}{12}$ & (1)     & 0.01 \\
\multicolumn{6}{c}{S-peak $(-60\arcsec, +60\arcsec)$} \\
{\soo}     &  3         &  19.7   & $\me{7.63}{14}$ & 1       & 0.04,1.21 \\
{\isoo}    &  2         &  (19.7) & $\me{5.63}{13}$ & (1)     & 0.04,0.05 \\
\hline
\multicolumn{6}{l}{A value surrounded by (...) indicates a fixed parameter.}
\end{tabular}
\end{table}

The derived {\hhco} rotation temperatures for the three cores are quite similar to
the ones obtained by \citet{2006A&A...453..949P} for other low-mass protostar
sources. Likewise, the derived {\hhco} column densities are just slightly higher
than the corresponding rotation diagram values of \citet{2006A&A...453..949P}. The
filling factor $\ebf=0.446$ determined toward the D-peak is in agreement with the
filling of $\sim 0.5$ determined from the {\ddco} $4_{0,4}-3_{0,3}$ source size in
\sref{s:hhco_res}.

\subsection{Analysis of the methanol lines}
\label{s:ch3oh_ana}
The {\chhhoh} analysis using the modified rotation diagram technique is based on
the integrated line intensities listed in \tref{t:ch3oh_intint}. In all sources 6
{\chhhoh} lines with lower state energies ranging from 23 to 46 K, have been used.
In the D-peak source a couple of lines are just marginally detected but are
included in the best fit analysis since they are weighted with the uncertainty and
do not affect the fit significantly. The line feature at 241904 MHz is a blend of
two {\chhhoh} lines of about the same energy and $A$-coefficient and we have not
been able to clearly resolve them into individual components. All lines at 241 GHz
were observed simultaneously. The only {\chhhoh} line not belonging to the 241 GHz
$5-4$ line forest is the $4_{2,2}-3_{1,2}$ E-line which is located in the 218 GHz
{\hhco}-band.

Only one line from the methanol A-species has been observed and included in the analysis.
Just like in the case of ortho and para formaldehyde, the A- and E-species of methanol are
treated together. The energy difference between the lowest A and E-methanol states is 8 K
with the A-species having the lowest energy. Furthermore, we only consider methanol to be
in its lowest torsional state.

The best fit results in the {\chhhoh} analysis have been entered in
\tref{t:h2co_rotdia}. In all three sources we find a low $\trot$ of 7 to 9 K.
Also the $\ebf$ is found to be very close to 1. The much lower rotation
temperatures found in the {\chhhoh} analysis as compared to the {\hhco} and
{\ddco} results suggest that the excitation of {\chhhoh} is quite sub-thermal and
a more elaborate treatment of the excitation and radiative transfer is needed
(as was demonstrated by \citet{1998A&A...335..266B} in their Fig. 9a).
This is also supported by the fact that none of the {\chhhoh} fits are good since
several of the modelled line intensities deviate substantially from the observed
line intensities. 

\subsection{ALI modelling}
\label{s:alimod}
Our aim with the non-LTE modelling using the ALI code is to construct a
homogeneous (in terms of physical parameters but not in terms of excitation
which may vary radially)
spherical model cloud, for each of the source positions (D-, P1, and
{\chhhoh}-peaks), that produces the observed {\hhco} and {\chhhoh} spectra. The
results from the modifed rotation diagram analysis above are used as a guide when
adopting the model cloud parameters. The ALI setup for the statistical equilibrium
equations includes collisional excitation rates and for formaldehyde we use the
He-{\hhco} collisional coefficients calculated by \citet{1991ApJS...76..979G}. The
coefficients have been multiplied by 2.2 to approximate the {\hh}-{\hhco}
collision system. For the {\hh}-{\chhhoh} collision coefficients we adopt those of
\citet{2004MNRAS.352...39P}. For both species only levels below 200 K are included
and we divide the model cloud into 29 shells. We do not include any radial
velocity field in the source and instead we simply use a microturbulent velocity
width, $\vt$, that reproduces the observed line widths. For an optically thin
line, the FWHM line width is $\sqrt{4\ln2}\,\vt$.

The results of the three ALI models have been summarized in \tref{t:alimodels}. The
listed cloud radii $R$ (in the first column) are based on the corresponding {\hhco} beam filling results
(\tref{t:h2co_rotdia}) and a distance to {\roa} of 120~pc. The tabulated column
densities are the source-averaged values $2\, N_\mathrm{peak}/3$, where
$N_\mathrm{peak}$ is the column density through the center of the spherical model
cloud. It should be noted that the derived $\nhh$ (and hence $N(\mathrm{H_2})$) is a
compromise value since, for all three models, a 20\% lower value yields better fits
for {\chhhoh} while a similarly higher value is optimum for {\hhco}. However, as the
differences are less than about 50\% any real difference in ${\nhh}$ is within the
uncertainties of the adopted collisional coefficients. The ortho-to-para ratio is
about 2 for all three models but since the ortho results are based on a single
optically thick line ($\tau\sim 4$) only and not observed simultaneously with the
three para lines, we cannot really exclude ortho-to-para ratios in the range
1-3. The derived {\hhco} column densities  are about 30\% lower than those derived
from the modified rotation diagram analysis. In contrast to $N(\mathrm{H_2CO})$ and
$N(\mathrm{H_2})$ which do not vary much among the three cores, the {\chhhoh} column
densities vary from $\me{5}{13}\cmcm$ in the D-peak to $\me{4}{14}\cmcm$ at the
{\chhhoh}-peak position. In addition, the E/A {\chhhoh} ratios seem to be close to 1
and since the A-line is observed simultaneously with the E-lines the determined
A/E-ratio of 1 is less uncertain than the estimated ortho/para ratios for
formaldehyde. The optical depth for the strongest of the {\chhhoh} lines are about
0.5-0.6 toward the {\chhhoh}-peak. The total {\hh} mass of a core is about
$0.1-0.2\,\mathrm{M}_\odot$.

\begin{table*}
\caption{Model cloud properties and results for the ALI analysis}
\label{t:alimodels}
\centering
\begin{tabular}{lrrrrrrrrrrrr}
\hline\hline
Source & $R/10^{16}$        & $\vt$   & $\tkin$  & $\nhh$  &   $N(\mathrm{H_2})$   & $M(\mathrm{H_2})$ &
\multicolumn{2}{c}{\Xmol{H_2CO}$\times 10^9$}  & $N(\mathrm{H_2CO})$ &
\multicolumn{2}{c}{\Xmol{CH_3OH}$\times 10^9$} & $N(\mathrm{CH_3OH})$ \\
       & (cm)      & ($\mathrm{km\, s^{-1}}$) & (K)      & ($\mathrm{cm^{-3}}$) & ($\mathrm{cm^{-2}}$) & ($\mathrm{M}_\odot$) &
  \mc{para}  &  \mc{ortho}        &  ($\mathrm{cm^{-2}}$) &  \mc{A} & \mc{E}  & ($\mathrm{cm^{-2}}$)  \\
\hline
D-peak         & 3.3 & 0.4 & 24 & $\me{6}{5}$  & $\me{3}{22}$ & 0.16 &
$1.8$ & $3.6$ & $\me{1.5}{14}$ &
$0.9$  & $0.9$ & $\me{5.0}{13}$ \\ 
P1   & 3.4 & 0.6 & 27 & $\me{7}{5}$  & $\me{3}{22}$ & 0.18 &
$1.7$  & $3.3$ & $\me{1.5}{14}$ &
$2.0$  & $2.0$ & $\me{1.2}{14}$ \\ 
{\chhhoh}-peak & 2.7 & 0.6 & 30 & $\me{1}{6}$  & $\me{4}{22}$ & 0.14 &
$1.7$  & $3.3$ & $\me{1.8}{14}$ &
$5.3$  & $5.8$ & $\me{4.0}{14}$ \\ 
\hline
\end{tabular}
\end{table*}

To our knowledge, there are no collision coefficients available for the collision
system {\hh}-{\ddco}, so we cannot use our ALI model for {\ddco}. However, using the
{\hh}-{\hhco} collision rates we check what happens to the excitation temperatures
when adopting a factor of 7 lower column density (the ratio of $N$(\hhco) and
$N$(\ddco) in \tref{t:h2co_rotdia} is close to 7). We then find that the excitation
temperatures of the $3_{1,2}-2_{1,1}$ transition drop between 20-30\% at different
radii. This drop in excitation temperature is in line with the 20\% lower
$\trot=17.4\K$ found for {\ddco} as compared to {\hhco} value of $\trot=22.5\K$
found toward the D-peak. Hence, it is quite plausible that the lower rotation
temperature found for {\ddco} as compared to {\hhco} can be explained by less
efficient photon trapping in the excitation of {\ddco}. We also checked the
influence on our results by continuum emission of dust, and the inclusion of a dust
component with standard dust parameters ($T_\mathrm{d}=22\K$, gas-to-dust mass ratio
of 100) had negligible impact on our modelling results.

\citet{2003A&A...402L..73L} report results from a large velocity gradient (LVG)
analysis of the {\roa} using {\chhhoh} $2-1$ and $3-2$ line forest data, at 96 and
145 GHz, obtained by SEST at an angular resolution larger ($52''$ and $35''$) than
for the present study. The analysed data were averaged from spectra spaced
by $30\arcsec$ in the north-south direction around the P2 position (see
\fref{f:maps}) and their spectra is thus a partial blend of the contribution
of the three cores. This
is especially the case for their $2-1$ observations. However, their LVG results
($\tkin = 20\K$, $\nhh =\me{4.5}{5}\iccm$, $X[\mathrm{CH_3OH}]=\me{2.7}{-9}$) are
in good agreement with those reported here.

\subsection{Analysis of other molecules}
\label{s:othana}
In the S-peak position we have detected three {\soo} lines and a modified
rotation diagram analysis gave best-fit results for $\ebf$ very close to 1. For
{\soo} we find a rotation temperature of about 20~K and a column density of
$N(\mathrm{SO_2})=\me{7.6}{14}\cmcm$. The same analysis was made for two of the
three detected {\isoo} lines and we get $N(\mathrm{^{34}SO_2})=\me{5.6}{13}\cmcm$
when using the rotation temperature of the more common variant. The
$11_{1,11}-10_{0,10}$ {\isoo} line was excluded in the fit since its observed
strength is incompatible (too strong by about a factor of 2) with the rotation
temperature found for {\soo}, cf. \tref{t:other_intint}. The reason for this is
unclear but could be due to an unknown blend or a non-LTE effect involving
$K_a=0$ states. Likewise, the two SO lines detected toward the D-peak position
result in a rotation temperature of 18~K and a column density of
$\me{2.4}{14}\cmcm$ (again assuming $\ebf=1$). In the P1 position, the two
detected {\soo} lines give $\trot = 23\K$ and a beam averaged column density of
$\me{9.9}{13}\cmcm$. Furthermore, assuming $\trot = 20\K$ we arrive at a beam
averaged {\soo} column density of $\me{8.3}{12}\cmcm$ toward the {\chhhoh}-peak.
The {\so} and {\soo} results have been entered in \tref{t:h2co_rotdia}.

The analysis results of the other molecules are summarized in \tref{t:othermodels}.
The column densities listed here are all calculated under the assumption of
optically thin emission. The {\cio}(3-2) column densities are based on the data from
\citet{2010A&A...510A..98L} and have been derived using the kinetic temperatures
derived from the ALI analysis as excitation temperature as the lower {\cio}
transitions are expected to be thermalized at the high densities of the cores. For
the S-peak, the {\soo} rotation temperature of 20~K was used as excitation
temperature for all molecules. Also tabulated is the $N$(\hh) column density deduced
from $N$(\cio) and assuming a {\cio} abundance of $X[\mathrm{C^{18}O}]=\me{2}{-7}$
(standard value of undepleted gas interstellar gas, \cite{1982ApJ...262..590F}, but
see also \cite{2005A&A...430..549W} for a discussion applicable to the $\rho$~Oph
cloud). In this context it is enlightning to estimate how much the cores contribute
to the observed {\cio}(3-2) integrated line intensity. Adopting the physical
parameters (\tref{t:alimodels}) and assuming $X[\mathrm{C^{18}O}]=\me{2}{-7}$ we
find, using the ALI code for {\cio}, that the cores make up 31\%, 37\%, and 67\% of
the observed {\cio}(3-2) emission in the D-, P1-, and {\chhhoh}-cores, respectively.
Firstly, this tells us that the {\cio} abundance in the cores cannot be much higher
than $\me{2}{-7}$ as they then would produce too much {\cio}(3-2) emission.
Secondly, an appreciable part of the observed {\cio}(3-2) emission is likely to
arise in a lower {\hh} density ($\la 10^5\iccm$) environment. This would explain the
higher $N$(\hh) column densities deduced from optically thin {\cio}
(\tref{t:othermodels}) as compared to the ALI results (\tref{t:alimodels}). A similar
finding was made by \citet{2005A&A...430..549W} in their CO study of other regions
in the $\rho$~Oph cloud where the cold ($\sim 10\K$) and dense ($\ga 10^5\iccm$)
cores were surrounded by a warmer ($\sim 30\K$) and less dense ($\sim 10^4\iccm$)
envelope. Also, given the observed ratio of {\cio}(3-2) and {\icio}(3-2) of about 23
toward the D-peak \citep{2010A&A...510A..98L}, it is likely that the {\cio}(3-2)
emission is somewhat optically thick ($\tau\approx 2$) and hence the listed
$N$(\cio) and $N$(\hh) (\tref{t:othermodels}) would be about a factor
$\tau/(1-e^{-\tau})\approx 2.3$ higher. The full {\hh} column density toward the
D-peak would then be $N(\mathrm{H_2}) = \me{1.5}{23}\cmcm$. Likewise, the {\so}
column density in the D-peak position of $\me{1.1}{14}\cmcm$ is lower than the
corresponding value of $\me{2.3}{14}\cmcm$  (\tref{t:h2co_rotdia}) where the latter
value includes compensation by optical depth. Using the ALI model for the D-peak
core (\tref{t:alimodels}) we find that adopting an abundance of
$X[\mathrm{SO}]=\me{1.6}{-8}$ results in integrated line intensites for the SO
$5_6-4_5$ and $6_5-5_4$ transitions that are close to the observed values
(\tref{t:other_intint}). Hence, using the previously derived physical
properties of the D-peak we can also explain the observed SO emission.
We assume that the bulk of SO emission originates in the
D-peak core itself and not from the low-density envelope. With the same
assumption for the optically thin {\soo} lines toward the D-peak, we find an
{\soo} abundance of $X[\mathrm{SO_2}] = N[\mathrm{SO_2}]/N[\mathrm{H_2}]
\approx (\me{4.23}{13}/0.446)/\me{3}{22} \approx \me{3}{-9}$ using the beam
averaged {\soo} column density of $\me{4.23}{13}\cmcm$, D-peak filling factor
of 0.446 (\tref{t:h2co_rotdia}) and the D-peak core {\hh} column density
(\tref{t:alimodels}). Similarly, for the P1 and {\chhhoh} peak positions we
obtain $X[\mathrm{SO_2}] = \me{7}{-9}$ and $\me{6}{-10}$, respectively. Here
the value for the P1 position is uncertain since it is likely that the {\soo}
emission do not originate in the same source as {\hhco} since they exhibit different
emission velocities (cf. \tref{t:peaks}). Using the P1 {\hh} column density
from {\cio}(3-2) (\tref{t:othermodels}) instead will lower the abundance by about
a factor of five. Taken together, the {\soo} abundances in the D-peak, P1, and
{\chhhoh} cores seem to fall in the range $\me{(0.6-3)}{-9}$. However, these
{\soo} abundances are all lower than the S-peak abundance, which
can be estimated to $X[\mathrm{SO_2}] = \me{2}{-8}$.

\begin{table}
\caption{Analysis results for other molecules assuming optically
thin emission}
\label{t:othermodels}
\centering
\begin{tabular}{lrrrr}
\hline\hline
$N$(\cio)	     & $N$(\hh)$^a$ & $N$(\iso) & $N$(\so) & $N$(\nndp) \\
($\mathrm{cm^{-2}}$) & ($\mathrm{cm^{-2}}$) &  ($\mathrm{cm^{-2}}$) &  ($\mathrm{cm^{-2}}$) &  ($\mathrm{cm^{-2}}$) \\
\hline
\multicolumn{5}{c}{D-peak $(0\arcsec,-30\arcsec)$} \\
$\me{1.3}{16}$       & $\me{6}{22}$ & $\me{6.4}{12}$    & $\me{1.1}{14}$  & $\me{2.3}{12}$ \\
\multicolumn{5}{c}{P1 $(0\arcsec,+30\arcsec)$} \\
$\me{1.4}{16}$       & $\me{7}{22}$ & $\me{1.6}{13}$    & $\me{2.0}{14}$  & $\me{2.6}{11}$ \\
\multicolumn{5}{c}{{\chhhoh}-peak $(-30\arcsec, 0\arcsec)$} \\
$\me{7.7}{15}$       & $\me{4}{22}$ & $<\!\me{2.2}{12}$ & $\me{1.2}{14}$  &  $\me{1.2}{11}$ \\
\multicolumn{5}{c}{S-peak $(-60\arcsec, +60\arcsec)^b$} \\
$\me{7.5}{15}$       & $\me{4}{22}$ & $\me{5.5}{13}$    & $\me{2.6}{14}$  &  $<\!\me{4.8}{10}$ \\
\hline
\multicolumn{4}{l}{$^a$ Assuming $X[\mathrm{C^{18}O}]=\me{2}{-7}$} \\
\multicolumn{4}{l}{$^b$ Adopting $\tex=20\,\mathrm{K}$} \\
\end{tabular}
\end{table}

\subsection{Column density ratios}
\label{s:colrat}
To estimate the ratios of different formaldehyde isotopologues we use the results
obtained by the modified rotation diagram technique (\tref{t:h2co_rotdia}). Even
for the main isotopic species we use the rotation diagram analysis results and not
the ALI results, since effects like adopting a spherical source will affect the
results and we rather be consequent in as many aspects as possible when estimating
column density ratios. The four column density ratios; {\hdco}/{\hhco},
{\ddco}/{\hdco}, {\ddco}/{\hhco}, {\hhco}/{\hhico} have been calculated for each
source position and entered into \tref{t:h2co_ratios}. The listed ratio
uncertainties have been estimated by including an absolute calibration uncertainty
of 10\% together with the $1\sigma$ uncertainty due to noise, where the latter is
the most prominent source of error for the weak lines. The {\hdco}/{\hhco} ratio of
$0.107\pm 0.015$ toward the D-peak is similar to the values found by
\citet{2006A&A...453..949P} for low-mass protostars. The {\ddco}/{\hdco} ratio of
$1.34\pm 0.19$ is higher than the \citet{2006A&A...453..949P} values which all fall
in the range 0.3-0.9. In fact, it is the only reported {\ddco}/{\hdco} ratio that is
significantly greater than 1.

\begin{table*}
\caption{Formaldehyde column density ratios}
\label{t:h2co_ratios}
\centering
\begin{tabular}{lrrrrrr}
\hline\hline
Source & {\hdco}/{\hhco} & {\ddco}/{\hdco} & {\ddco}/{\hhco} & {\hhco}/{\hhico} & $F$  & {\hhco}/{\chhhoh} \\
\hline
D-peak         $(0\arcsec,-30\arcsec)$  & $0.107\pm 0.015$ & $1.34\pm 0.19$ & $0.143\pm 0.020$   & $54\pm 13$  & $0.080\pm 0.016$ & $3.0\pm 0.4$ \\ 
P1             $(0\arcsec,+30\arcsec)$  & $0.033\pm 0.006$ & $0.27\pm 0.08$ & $0.0087\pm 0.0024$ & $111\pm 32$ & $0.12\pm 0.04$ & $1.2\pm 0.2$ \\ 
{\chhhoh}-peak $(-30\arcsec, 0\arcsec)$ & $0.009\pm 0.004$ & $0.8\pm 0.5$   & $0.0073\pm 0.0025$ & $74\pm 18$  &  & $0.45\pm 0.06$ \\ 
\hline
\end{tabular}
\end{table*}

The quantity
\begin{equation}
 F = \frac{(\mathrm{HDCO}/\mathrm{H_2CO})^2}{\mathrm{D_2CO}/\mathrm{H_2CO}} =
     \frac{\mathrm{HDCO}/\mathrm{H_2CO}}{\mathrm{D_2CO}/\mathrm{HDCO}}
\label{e:F}
\end{equation}
is also tabulated in \tref{t:h2co_ratios}. This quantity was introduced
by \citet{2002P&SS...50.1125R} to discern between gas phase and grain surface formation
mechanisms of deuterated formaldehyde. Our D-peak $F$-value of 0.08 is
similar to the values 0.03-0.2 found by \citet{2007A&A...471..849R}. In the last
column of \tref{t:h2co_ratios} we also list the {\hhco}/{\chhhoh} column
density ratios based on the ALI model results.

\citet{2001ApJ...562L.185B} used observations of rare {\co} isotopologues to
determine the $\mathrm{^{12}C/^{13}C}$ ratio toward {\roc} to $65\pm 11$. Although
the {\roc} cloud core is quite far from the {\roa} position they belong to the same
nearby cloud complex and our {\hhco}/{\hhico} ratios (\tref{t:h2co_ratios}) are in
good agreement with their value. Consequently, we will adopt this
$\mathrm{^{12}C/^{13}C}$ ratio of $65$ also for {\roa} and the opacity for the
{\cio}(3-2) line toward the D-peak can be estimated from the optical depth of the
{\icio}(3-2) line of 0.03 \citep{2010A&A...510A..98L} as
$\tau[\mathrm{C^{18}O(3-2)}] \approx 65\times 0.03 \approx 2$. This is then
consistent with the discussion in the previous section on the {\cio}(3-2) opacity
corrections toward the D-peak. 

In \tref{t:other_ratios} the {\so}/{\iso} and {\soo}/{\isoo} column density ratios
have been listed. As noted in the table, some of the ratios are affected by opacity
in the main isotopologue and should be regarded as lower limits.  The interstellar
$\mathrm{^{32}S}/\mathrm{^{34}S}$ ratio \citep{1998A&A...337..246L} appears to be
close to its solar value of 22 \citep{2009ARA&A..47..481A}. We can use this
isotopologue ratio to estimate the SO abundances from the expression $22\,
N[\mathrm{^{34}SO}]/N[\mathrm{H_2}]$. For the D-peak we already have an abundance
estimate, made in the previous section, of $X[\mathrm{SO}]=\me{1.6}{-8}$. In the P1
and {\chhhoh} peaks we then get $\me{2}{-8}$ and $\me{3}{-9}$, respectively. The P1
value is quite uncertain, as the {\so} emission velocities differ here from that of
{\hhco} (cf. the case of {\soo} in the previous section), will be lower by a factor
5 when using beam averaged column densities. Using the values in
\tref{t:othermodels} for the S-peak we find that the {\so} abundance is
$X[\mathrm{SO}] = 22\times\me{5.5}{13}/\me{4}{22} \approx \me{3}{-8}$.

\begin{table}
\caption{Other column density ratios}
\label{t:other_ratios}
\centering
\begin{tabular}{lrrr}
\hline\hline
Source & {\so}/{\iso} & {\soo}/{\isoo} & {\nndp}/{\nnhp}$^{ab}$ \\
\hline
D-peak         & $25\pm 5$      &           & $0.078\pm 0.011$	 \\ 
P1             & $13\pm 2^b$    &           & $0.0083\pm 0.0024$  \\ 
{\chhhoh}-peak & $> 18^c$       &           & $0.024\pm 0.010$  \\ 
S-peak         & $4.7\pm 0.7^b$ & $14\pm 3$ &   \\ 
\hline
\multicolumn{4}{l}{$^a$ $N$(\nnhp) estimated from \citet{2007A&A...472..519A}} \\
\multicolumn{4}{l}{$^b$ Not corrected for opacity} \\
\multicolumn{4}{l}{$^c$ 3$\sigma$ limit}
\end{tabular}
\end{table}

The ratio {\nndp}/{\nnhp} is also tabulated in \tref{t:other_ratios} and here we
use the {\nnhp}(1-0) integrated intensity data from \citet{2007A&A...472..519A}
which have a beam size of $26\arcsec$, i.e. almost identical to the resolution of
our {\nndp}(3-2) data. The listed $N$({\nndp})/$N$({\nnhp}) ratios have not been
compensated for optical depth of the {\nnhp} lines.
 
\section{Discussion}
\label{s:dis}

\subsection{Interaction with the molecular outflow?}
\label{s:outflow}
As described in \sref{s:res_oth}, the observed line profiles for {\so} and {\soo}
show broader wings in the north-western corner of our map (\fref{f:so_34so}).
Also the {\hhco} spectra exhibit line wings in this region (\fref{f:h2co_d2co}).
The CO red wing emission from the prominent outflow emanating
from VLA~1623 extends in this direction while the blue wing emission is extended
both in the NW and SE directions \citep{1990A&A...236..180A}. The {\so} and {\soo}
emission peaks very close to the IR reflection nebula GSS30
\citep{1985ApJ...290..261C} which is at offsets $(-79\arcsec,+28\arcsec)$ in our
maps. The {\so} and {\soo} abundances ($\me{3}{-8}$ and $\me{2}{-8}$,
respectively) are here clearly higher than in other positions. The {\so} and {\soo} line
wings could very well be associated with GSS30 rather than the VLA~1623 CO
outflow. The {\chhhoh}-peak at $(-30\arcsec, 0\arcsec)$, on the other hand, could
be a result of an interaction with the outflow and a dense clump
\citep[cf.][]{2004A&A...415.1021J}. The methanol abundance is about 5 times higher
here compared to its abundance in the D-peak and P1 cores. It appears also to be
somewhat warmer ($\sim 30\K$) than the other cores.

\subsection{Depletion}
\label{s:depletion}
\begin{figure}
\includegraphics[width=9cm]{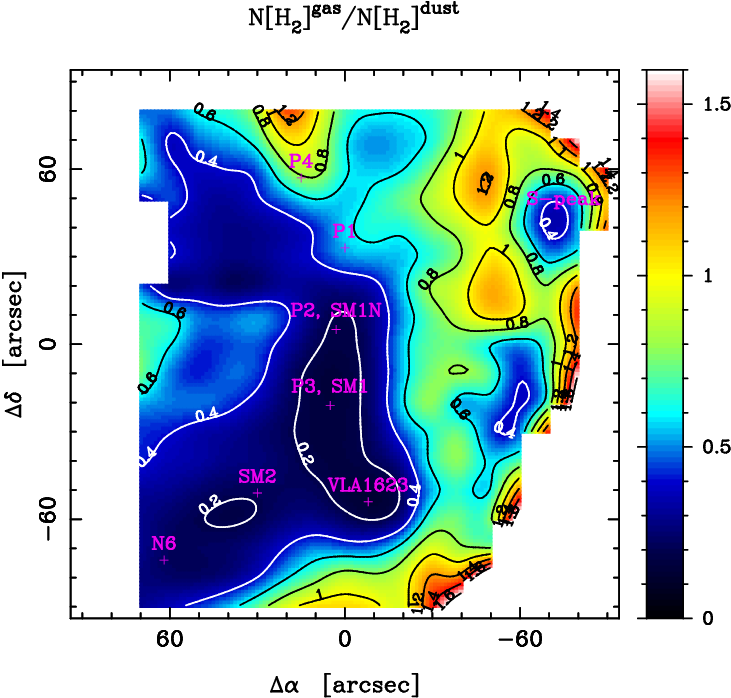}
\caption{Gas-to-dust ratio map of {\roa} calculated as the {\hh} column density from
{\cio} relative to the {\hh} column density from dust. The gas column data are based on
the {\cio}(3-2) map of \citet{2010A&A...510A..98L} and the dust continuum data are from
\cite{1998A&A...336..150M}. The gas column data have not been corrected for opacity.}
\label{f:ratio_map}
\end{figure}

The D-peak position where the deuterium fractionation is highest coincides with the
emission peak (SM1) in the 1.3 mm continuum  map of \citet{1998A&A...336..150M}. The 1.3
mm flux density, $S_\mathrm{1.3mm}$, can be converted into an {\hh} column density using
$\Nhh = \me{4}{20}\cmcm (S_\mathrm{1.3mm}/\mathrm{mJy\, beam^{-1}})$ assuming
$T_\mathrm{dust}=20\K$, a dust mass opacity of $\kappa_\mathrm{1.3mm} = 0.5\cmcmg$,
and a gas-to-dust mass ratio of 100 \citep[e.g.][]{1996A&A...314..625A} over the
entire map \citep[see Eq. ($1'$) in][]{1998A&A...336..150M}. This can be compared to the
{\hh} column density derived from the {\cio}(3-2) map of \citet{2010A&A...510A..98L}.
Following the discussion in \sref{s:othana} we here assume
$X[\mathrm{C^{18}O}]=\me{2}{-7}$ and an excitation temperature of $20\K$ in all
positions. The ratio, $\Nhh^\mathrm{gas}/\Nhh^\mathrm{dust}$, of these two estimates of
$\Nhh$ are displayed in \fref{f:ratio_map}. The lowest values of this ratio are around
0.15 and they are found in the central part of the cloud. Opacity corrections for {\cio}
in this central part are of the order 2 (which will increase the ratio to 0.3), see
\sref{s:othana} and probably less (closer to 1) further out. This would still mean that
the $\Nhh^\mathrm{gas}$ estimate is a factor of 3 lower than that from dust in the
central portions of {\roa}. Here the major uncertainty is likely to be related to the
adopted {\cio} abundance and dust mass opacity, $\kappa_\mathrm{1.3mm}$, while
influences from temperature changes are smaller. The low ratios seen could be due to
depletion of CO and would then correspond to a CO depletion factor of 2-3. Given the
uncertainty of the adopted dust parameters and degree of depletion (ie. {\cio}
abundance) we estimate that the full {\hh} column density toward the D-peak, cf.
\sref{s:othana}, is in the range $\me{(1-4)}{23}\cmcm$.

The freeze-out of CO on grains (together with low temperatures) is thought to
increase the deuterium fractionation. Although the D-peak at $(0\arcsec,-30\arcsec)$
is located in the part where the $\Nhh^\mathrm{gas}/\Nhh^\mathrm{dust}$  opacity
corrected ratio is low $\sim 0.3-0.4$, we do not see any signs that CO depletion is
in effect more at $(0\arcsec,-30\arcsec)$ than in any of the other central
positions. Hence, the amount of CO depletion (at least as measured by the {\cio}
emission and the 1.3 mm continuum data) is not directly responsible to the elevated
deuterium fractionation seen in formaldehyde toward the D-peak. However, the D-peak
is located within a larger region that is likely to be depleted and this may have
been important in an earlier stage of the evolution of the D-peak source.

Interestingly, there is a minimum in the $\Nhh^\mathrm{gas}/\Nhh^\mathrm{dust}$
ratio in the SE part of the map (\fref{f:ratio_map}). This minimum coincides
with the secondary {\nndp} peak (N6). This would be in line with the findings that
the amount of CO depletion and the {\nndp}/{\nnhp} ratio are correlated in starless
cores \citep{2005ApJ...619..379C}. Moreover, the Class 0 source VLA~1623 is also
associated with a minimum in the $\Nhh^\mathrm{gas}/\Nhh^\mathrm{dust}$ ratio and
possibly also with a higher {\nndp}/{\nnhp} ratio, compare \fref{f:nndp_map} and
\fref{f:ratio_map}. A correlation between CO depletion and {\nndp}/{\nnhp} ratio has
been found by \citet{2009A&A...493...89E} toward several Class 0 objects. 

\subsection{Gas-phase chemistry models}
\label{s:gas_phase}
\begin{figure}
\includegraphics[width=9cm]{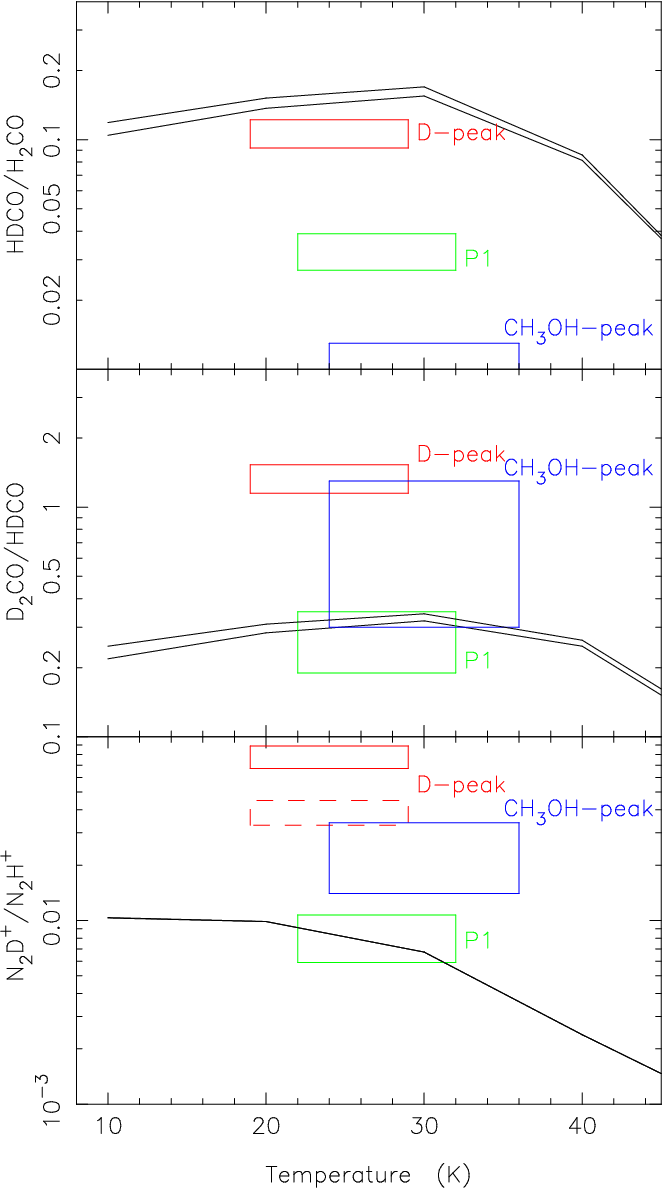}
\caption{Gas-phase steady-state models based on the reaction network by
\citet{2007A&A...464..245R}. Shown are the D/H predicted ratios; {\hdco}/{\hhco} (top),
{\ddco}/{\hdco} (middle), and {\nndp}/{\nnhp} (bottom) as function of temperature and
for two different {\hh} densities ($\me{5}{5}\iccm$ and $\me{1}{6}\iccm$). The coloured
boxes correspond to the observed ratios for the three cores as indicated in the top
panel. The size of the rectangles represents the estimated uncertainties.}
\label{f:gas_phase}
\end{figure}

Pure gas-phase chemistry can lead to elevated deuterium fractionations for
{\hhco}. We have here adopted the low-metalicity (or depleted) model of
\citet{2007A&A...464..245R} as their warm-core (and un-depleted) model shows low
deuterium fractionations. The steady-state model work presented here includes the
updates and modifications of \citet{2009A&A...508..737P}, and we have used it to
predict relevant D/H ratios for {\hhco} and {\nnhp}. The models are shown in
\fref{f:gas_phase}. The different steady-state D/H model ratios are shown as a
function of temperature and have been calculated for two different {\hh}
densities; $\me{5}{5}\iccm$ and $\me{1}{6}\iccm$. The modelled ratios show little
variation due to the used {\hh} density. The observed column density ratios (taken
from \tref{t:h2co_ratios} and \tref{t:other_ratios}) are represented as coloured
boxes with size according to estimated uncertainty. The dashed rectangle indicates
the {\nndp}/{\nnhp} ratio toward the D-peak with an opacity correction.

It is noteworthy that the gas-phase depleted models yield $F$-values, see
\eref{e:F}, very close to 0.5 for $T\la 30\K$ for formaldehyde
which are significantly higher than the observed values
$F\approx 0.1$, cf. \tref{t:h2co_ratios}. The absolute {\hhco} abundances,
$\me{(3-6)}{-11}$, for these models are much lower than the observed {\hhco}
abundances $\sim \me{5}{-9}$.

\subsection{Grain chemistry models}
\label{s:grain}
Our observations point to the fact that {\ddco} is more abundant than
{\hdco} toward the D-peak. This is, to our knowledge, the first source
where such case is observed. We note that {\hdco} and {\hhco} showed a
similar anomaly towards L1527 \citep{2006A&A...453..949P}, but even in that
case, {\ddco} was less abundant than {\hdco}. This anomaly is difficult to
explain in terms of pure gas-phase chemistry, as discussed above.

We discuss here the possibility that this anomaly stems from grain chemistry.
Simple grain models which only account for H and D additions to CO to form
formaldehyde and methanol cannot account for enrichment of {\ddco} compared to the
statistically expected value. However, including the abstraction or exchange
reactions of the type $\mathrm{HDCO + H \to DCO + H_2}$,
\citet{2002P&SS...50.1125R} argue that the enrichment is anyway limited to
{\hdco}/{\hhco} $>$ {\ddco}/{\hdco} (or $F > 1$, see \eref{e:F}). Our observational results (with
$F\approx 0.1$), as well as those of \citet{2006A&A...453..949P} and
\citet{2007A&A...471..849R}, are contradictory to this model prediction, showing
that even this detailed model may be lacking some processes.

Some laboratory experiments have aimed in the last years to explain the large
fractionations of formaldehyde and methanol observed towards low-mass protostars.
Different types of experiments have been done \citep[see][for an
overview]{2008PrSS...83..439W}. \citet{2005ApJ...624L..29N} have exposed CO ices
with H and D atoms (with D/H of 0.1) and have shown that {\hhco} and {\chhhoh} (as
well as their deuterated counterparts) are formed. In this experiment, they can
reproduce the fractionation of all deuterated methanol molecules observed towards
IRAS16293$-$2442 \citep{2004A&A...416..159P}. However, they never observe an
enrichment of {\ddco} higher than {\hdco} (Naoki Watanabe, priv. comm.). In a
second type of experiment, \citet{2009ApJ...702..291H} aimed at clarifying the
different formation routes by exposing  an {\hhco} ice sample with D atoms.
Exchanges are observed, showing that deuterium fractionation occurs very
efficiently along the reaction path $\mathrm{H_2CO \to HDCO \to D_2CO}$. In this
case, the observed {\ddco} can become more abundant than {\hdco} (see their Figure
3). Although a direct extrapolation from these experiments is difficult, our
observations may be the definitive evidence that abstraction and exchange
reactions are playing an important role in grain chemistry. It is not clear if
reproducing our observations will require an increased atomic D/H ratio
\citep[from the 0.1 value used by][]{2005ApJ...624L..29N} incoming on the grains,
or if the longer timescales involved in the ISM chemistry compared to the
laboratory experiments also can play a role. Detailed modelling of the complex
processes taking place on the grains would be needed to definitely settle this
question.

There is no known central source towards the D-peak. So, unlike the case of
IRAS16293$-$2422, there is no protostar that can be responsible for releasing the
deuterated material into the gas. We therefore
need to invoke heating of the grains by cosmic rays \citep[e.g.][]{2006PNAS..10312257H}
or through the release of the
formation energy of the newly formed species \citep{2006FaDi..133...51G}.
Both these mechanisms can be of importance for desorption of grain mantles in
starless cores. However, even if these mechanisms are responsible of
releasing the deuterated material into the gas, there is no obvious reason why they
would be more effective in the SM1 core as compared to the other cores. Hence,
the underlying question why the D-peak shows such a high deuterium fractionation
remains unanswered.

\section{Conclusions}
\label{s:con}
We have observed a very high degree of deuteration of formaldehyde towards a core
(SM1) in {\roa}. In this D-peak, the deuterium fractionation manifests itself by a
very high {\ddco}/{\hdco} ratio of $1.34\pm 0.19$ while the ratio {\hdco}/{\hhco} is
$0.107\pm 0.015$. In the other parts of this cloud core the degree of deuteration in
formaldehyde is much lower (although still very high compared to the ISM D/H ratio).
For instance, at the P1 position, about $1'$ (or 0.035 pc) north of the D-peak, the
corresponding ratios are $0.27\pm 0.08$ and $0.033\pm 0.006$, respectively. A similar
decrease in deuterium fractionation between the two positions is also seen for
{\nndp}/{\nnhp}. The {\hhco} abundance relative
to {\hh} is estimated to be around $\me{5}{-9}$ over the central core.

The {\chhhoh} distribution is clearly different from that of {\hhco} and it has its
maximum about $45''$ to the northwest of the deuterium peak. The elevated methanol
abundance (by about a factor of 5 relative to the D-peak) here could be due to an
interaction of the outflow with a dense clump. It would be interesting to study how
the deuterated versions of {\chhhoh} are distributed in this source
\citep[cf.][]{2006A&A...453..949P}.

In order to understand the reason of the very high deuteration level observed toward
the D-peak we have performed gas-phase chemistry modelling (for a depleted source).
By looking at the ratio ({\hdco}/{\hhco})/({\ddco}/{\hdco}) (the $F$-value, see \eref{e:F}) we find
that the models result in values around 0.5 while the observed values are always
around 0.1. Also, the absolute {\hhco} abundances obtained in these models are too
low by a factor of 100. Hence, the gas-phase chemistry scheme cannot account for the
observed {\hhco} abundances and deuterium ratios. Instead, we advocate that grain
chemistry, in terms of abstraction and exchange reactions in the reaction chain
$\mathrm{H_2CO \to HDCO \to D_2CO}$ \citep{2009ApJ...702..291H}, can be responsible
for the very high deuterium fractionations observed in {\roa}. However, before being
too conclusive about, e.g., the required atomic D/H ratio, these grain chemistry model
results need to be expressed in observable quantities like the {\hhco} $F$-value.
Again, observations of multiply deuterated methanol isotopologues toward the D-peak
in {\roa} will provide additional insight as they are key ingredients in the
grain chemistry scheme. 

\begin{acknowledgements}
  We are very grateful to Fr{\'e}d{\'e}rique Motte for sending us her continuum
  data of $\rho$~Oph and to Naoki Watanabe for sending us unpublished laboratory
  results.
  Excellent support from the APEX staff during the observations is also greatly
  appreciated. BP is funded by the Deutsche Forschungsgemeinschaft (DFG) under the
  Emmy Noether project number PA1692/1-1.
\end{acknowledgements}
\bibliographystyle{aa} 
\bibliography{15012} 

\begin{thebibliography}{70}
\expandafter\ifx\csname natexlab\endcsname\relax\def\natexlab#1{#1}\fi

\bibitem[{{Andr{\'e}} {et~al.}(2007){Andr{\'e}}, {Belloche}, {Motte}, \&
  {Peretto}}]{2007A&A...472..519A}
{Andr{\'e}}, P., {Belloche}, A., {Motte}, F., \& {Peretto}, N. 2007, \aap, 472,
  519

\bibitem[{{Andr{\'e}} {et~al.}(1990){Andr{\'e}}, {Martin-Pintado}, {Despois},
  \& {Montmerle}}]{1990A&A...236..180A}
{Andr{\'e}}, P., {Martin-Pintado}, J., {Despois}, D., \& {Montmerle}, T. 1990,
  \aap, 236, 180

\bibitem[{{Andr{\'e}} {et~al.}(1993){Andr{\'e}}, {Ward-Thompson}, \&
  {Barsony}}]{1993ApJ...406..122A}
{Andr{\'e}}, P., {Ward-Thompson}, D., \& {Barsony}, M. 1993, \apj, 406, 122

\bibitem[{{Andre} {et~al.}(1996){Andre}, {Ward-Thompson}, \&
  {Motte}}]{1996A&A...314..625A}
{Andre}, P., {Ward-Thompson}, D., \& {Motte}, F. 1996, \aap, 314, 625

\bibitem[{{Asplund} {et~al.}(2009){Asplund}, {Grevesse}, {Sauval}, \&
  {Scott}}]{2009ARA&A..47..481A}
{Asplund}, M., {Grevesse}, N., {Sauval}, A.~J., \& {Scott}, P. 2009, \araa, 47,
  481

\bibitem[{{Asvany} {et~al.}(2004){Asvany}, {Schlemmer}, \&
  {Gerlich}}]{2004ApJ...617..685A}
{Asvany}, O., {Schlemmer}, S., \& {Gerlich}, D. 2004, \apj, 617, 685

\bibitem[{{Bachiller} {et~al.}(1998){Bachiller}, {Codella}, {Colomer},
  {Liechti}, \& {Walmsley}}]{1998A&A...335..266B}
{Bachiller}, R., {Codella}, C., {Colomer}, F., {Liechti}, S., \& {Walmsley},
  C.~M. 1998, \aap, 335, 266

\bibitem[{{Bacmann} {et~al.}(2003){Bacmann}, {Lefloch}, {Ceccarelli},
  {Steinacker}, {Castets}, \& {Loinard}}]{2003ApJ...585L..55B}
{Bacmann}, A., {Lefloch}, B., {Ceccarelli}, C., {et~al.} 2003, \apjl, 585, L55

\bibitem[{{Bensch} {et~al.}(2001){Bensch}, {Pak}, {Wouterloot}, {Klapper}, \&
  {Winnewisser}}]{2001ApJ...562L.185B}
{Bensch}, F., {Pak}, I., {Wouterloot}, J.~G.~A., {Klapper}, G., \&
  {Winnewisser}, G. 2001, \apjl, 562, L185

\bibitem[{{Castelaz} {et~al.}(1985){Castelaz}, {Hackwell}, {Grasdalen},
  {Gehrz}, \& {Gullixson}}]{1985ApJ...290..261C}
{Castelaz}, M.~W., {Hackwell}, J.~A., {Grasdalen}, G.~L., {Gehrz}, R.~D., \&
  {Gullixson}, C. 1985, \apj, 290, 261

\bibitem[{{Ceccarelli} {et~al.}(1998){Ceccarelli}, {Castets}, {Loinard},
  {Caux}, \& {Tielens}}]{1998A&A...338L..43C}
{Ceccarelli}, C., {Castets}, A., {Loinard}, L., {Caux}, E., \& {Tielens},
  A.~G.~G.~M. 1998, \aap, 338, L43

\bibitem[{{Ceccarelli} {et~al.}(2001){Ceccarelli}, {Loinard}, {Castets},
  {Tielens}, {Caux}, {Lefloch}, \& {Vastel}}]{2001A&A...372..998C}
{Ceccarelli}, C., {Loinard}, L., {Castets}, A., {et~al.} 2001, \aap, 372, 998

\bibitem[{{Ceccarelli} {et~al.}(2002){Ceccarelli}, {Vastel}, {Tielens},
  {Castets}, {Boogert}, {Loinard}, \& {Caux}}]{2002A&A...381L..17C}
{Ceccarelli}, C., {Vastel}, C., {Tielens}, A.~G.~G.~M., {et~al.} 2002, \aap,
  381, L17

\bibitem[{{Crapsi} {et~al.}(2005){Crapsi}, {Caselli}, {Walmsley}, {Myers},
  {Tafalla}, {Lee}, \& {Bourke}}]{2005ApJ...619..379C}
{Crapsi}, A., {Caselli}, P., {Walmsley}, C.~M., {et~al.} 2005, \apj, 619, 379

\bibitem[{{Di Francesco} {et~al.}(2004){Di Francesco}, {Andr{\'e}}, \&
  {Myers}}]{2004ApJ...617..425D}
{Di Francesco}, J., {Andr{\'e}}, P., \& {Myers}, P.~C. 2004, \apj, 617, 425

\bibitem[{{Di Francesco} {et~al.}(2009){Di Francesco}, {Andr{\'e}}, \&
  {Myers}}]{2009ApJ...700.1994D}
{Di Francesco}, J., {Andr{\'e}}, P., \& {Myers}, P.~C. 2009, \apj, 700, 1994

\bibitem[{{Emprechtinger} {et~al.}(2009){Emprechtinger}, {Caselli}, {Volgenau},
  {Stutzki}, \& {Wiedner}}]{2009A&A...493...89E}
{Emprechtinger}, M., {Caselli}, P., {Volgenau}, N.~H., {Stutzki}, J., \&
  {Wiedner}, M.~C. 2009, \aap, 493, 89

\bibitem[{{Frerking} {et~al.}(1982){Frerking}, {Langer}, \&
  {Wilson}}]{1982ApJ...262..590F}
{Frerking}, M.~A., {Langer}, W.~D., \& {Wilson}, R.~W. 1982, \apj, 262, 590

\bibitem[{{Garrod} {et~al.}(2006){Garrod}, {Park}, {Caselli}, \&
  {Herbst}}]{2006FaDi..133...51G}
{Garrod}, R., {Park}, I.~H., {Caselli}, P., \& {Herbst}, E. 2006, Chemical
  Evolution of the Universe, Faraday Discussions, volume 133, 2006, p.51, 133,
  51

\bibitem[{{Gerlich} {et~al.}(2002){Gerlich}, {Herbst}, \&
  {Roueff}}]{2002P&SS...50.1275G}
{Gerlich}, D., {Herbst}, E., \& {Roueff}, E. 2002, \planss, 50, 1275

\bibitem[{{Goldsmith} \& {Langer}(1999)}]{1999ApJ...517..209G}
{Goldsmith}, P.~F. \& {Langer}, W.~D. 1999, \apj, 517, 209

\bibitem[{{Green}(1991)}]{1991ApJS...76..979G}
{Green}, S. 1991, \apjs, 76, 979

\bibitem[{{G{\"u}sten} {et~al.}(2006){G{\"u}sten}, {Nyman}, {Schilke},
  {Menten}, {Cesarsky}, \& {Booth}}]{2006A&A...454L..13G}
{G{\"u}sten}, R., {Nyman}, L.~{\AA}., {Schilke}, P., {et~al.} 2006, \aap, 454,
  L13

\bibitem[{{Herbst} {et~al.}(1987){Herbst}, {Adams}, {Smith}, \&
  {Defrees}}]{1987ApJ...312..351H}
{Herbst}, E., {Adams}, N.~G., {Smith}, D., \& {Defrees}, D.~J. 1987, \apj, 312,
  351

\bibitem[{{Herbst} \& {Cuppen}(2006)}]{2006PNAS..10312257H}
{Herbst}, E. \& {Cuppen}, H.~M. 2006, Proceedings of the National Academy of
  Science, 103, 12257

\bibitem[{{Hidaka} {et~al.}(2009){Hidaka}, {Watanabe}, {Kouchi}, \&
  {Watanabe}}]{2009ApJ...702..291H}
{Hidaka}, H., {Watanabe}, M., {Kouchi}, A., \& {Watanabe}, N. 2009, \apj, 702,
  291

\bibitem[{{Jayawardhana} {et~al.}(2001){Jayawardhana}, {Hartmann}, \&
  {Calvet}}]{2001ApJ...548..310J}
{Jayawardhana}, R., {Hartmann}, L., \& {Calvet}, N. 2001, \apj, 548, 310

\bibitem[{{J{\o}rgensen} {et~al.}(2004){J{\o}rgensen}, {Hogerheijde}, {Blake},
  {van Dishoeck}, {Mundy}, \& {Sch{\"o}ier}}]{2004A&A...415.1021J}
{J{\o}rgensen}, J.~K., {Hogerheijde}, M.~R., {Blake}, G.~A., {et~al.} 2004,
  \aap, 415, 1021

\bibitem[{{Klein} {et~al.}(2006){Klein}, {Philipp}, {Kr{\"a}mer}, {Kasemann},
  {G{\"u}sten}, \& {Menten}}]{2006A&A...454L..29K}
{Klein}, B., {Philipp}, S.~D., {Kr{\"a}mer}, I., {et~al.} 2006, \aap, 454, L29

\bibitem[{{Larsson} {et~al.}(2007){Larsson}, {Liseau}, {Pagani}, {Bergman},
  {Bernath}, {Biver}, {Black}, {Booth}, {Buat}, {Crovisier}, {Curry},
  {Dahlgren}, {Encrenaz}, {Falgarone}, {Feldman}, {Fich}, {Flor{\'e}n},
  {Fredrixon}, {Frisk}, {Gahm}, {Gerin}, {Hagstr{\"o}m}, {Harju}, {Hasegawa},
  {Hjalmarson}, {Johansson}, {Justtanont}, {Klotz}, {Kyr{\"o}l{\"a}}, {Kwok},
  {Lecacheux}, {Liljestr{\"o}m}, {Llewellyn}, {Lundin}, {M{\'e}gie},
  {Mitchell}, {Murtagh}, {Nordh}, {Nyman}, {Olberg}, {Olofsson}, {Olofsson},
  {Olofsson}, {Persson}, {Plume}, {Rickman}, {Ristorcelli}, {Rydbeck},
  {Sandqvist}, {Sch{\'e}ele}, {Serra}, {Torchinsky}, {Tothill}, {Volk},
  {Wiklind}, {Wilson}, {Winnberg}, \& {Witt}}]{2007A&A...466..999L}
{Larsson}, B., {Liseau}, R., {Pagani}, L., {et~al.} 2007, \aap, 466, 999

\bibitem[{{Linsky}(2003)}]{2003SSRv..106...49L}
{Linsky}, J.~L. 2003, Space Science Reviews, 106, 49

\bibitem[{{Lis} {et~al.}(2002){Lis}, {Roueff}, {Gerin}, {Phillips}, {Coudert},
  {van der Tak}, \& {Schilke}}]{2002ApJ...571L..55L}
{Lis}, D.~C., {Roueff}, E., {Gerin}, M., {et~al.} 2002, \apjl, 571, L55

\bibitem[{{Liseau} {et~al.}(2010){Liseau}, {Larsson}, {Bergman}, {Pagani},
  {Black}, {Hjalmarson}, \& {Justtanont}}]{2010A&A...510A..98L}
{Liseau}, R., {Larsson}, B., {Bergman}, P., {et~al.} 2010, \aap, 510, A98+

\bibitem[{{Liseau} {et~al.}(2003){Liseau}, {Larsson}, {Brandeker}, {Bergman},
  {Bernath}, {Black}, {Booth}, {Buat}, {Curry}, {Encrenaz}, {Falgarone},
  {Feldman}, {Fich}, {Flor{\'e}n}, {Frisk}, {Gerin}, {Gregersen}, {Harju},
  {Hasegawa}, {Hjalmarson}, {Johansson}, {Kwok}, {Lecacheux}, {Liljestr{\"o}m},
  {Mattila}, {Mitchell}, {Nordh}, {Olberg}, {Olofsson}, {Pagani}, {Plume},
  {Ristorcelli}, {Sandqvist}, {Sch{\'e}ele}, {Serra}, {Tothill}, {Volk}, \&
  {Wilson}}]{2003A&A...402L..73L}
{Liseau}, R., {Larsson}, B., {Brandeker}, A., {et~al.} 2003, \aap, 402, L73

\bibitem[{{Loinard} {et~al.}(2001){Loinard}, {Castets}, {Ceccarelli}, {Caux},
  \& {Tielens}}]{2001ApJ...552L.163L}
{Loinard}, L., {Castets}, A., {Ceccarelli}, C., {Caux}, E., \& {Tielens},
  A.~G.~G.~M. 2001, \apjl, 552, L163

\bibitem[{{Loinard} {et~al.}(2002){Loinard}, {Castets}, {Ceccarelli},
  {Lefloch}, {Benayoun}, {Caux}, {Vastel}, {Dartois}, \&
  {Tielens}}]{2002P&SS...50.1205L}
{Loinard}, L., {Castets}, A., {Ceccarelli}, C., {et~al.} 2002, \planss, 50,
  1205

\bibitem[{{Loinard} {et~al.}(2000){Loinard}, {Castets}, {Ceccarelli},
  {Tielens}, {Faure}, {Caux}, \& {Duvert}}]{2000A&A...359.1169L}
{Loinard}, L., {Castets}, A., {Ceccarelli}, C., {et~al.} 2000, \aap, 359, 1169

\bibitem[{{Loinard} {et~al.}(2008){Loinard}, {Torres}, {Mioduszewski}, \&
  {Rodr{\'{\i}}guez}}]{2008ApJ...675L..29L}
{Loinard}, L., {Torres}, R.~M., {Mioduszewski}, A.~J., \& {Rodr{\'{\i}}guez},
  L.~F. 2008, \apjl, 675, L29

\bibitem[{{Lombardi} {et~al.}(2008){Lombardi}, {Lada}, \&
  {Alves}}]{2008A&A...480..785L}
{Lombardi}, M., {Lada}, C.~J., \& {Alves}, J. 2008, \aap, 480, 785

\bibitem[{{Loren} {et~al.}(1990){Loren}, {Wootten}, \&
  {Wilking}}]{1990ApJ...365..269L}
{Loren}, R.~B., {Wootten}, A., \& {Wilking}, B.~A. 1990, \apj, 365, 269

\bibitem[{{Lucas} \& {Liszt}(1998)}]{1998A&A...337..246L}
{Lucas}, R. \& {Liszt}, H. 1998, \aap, 337, 246

\bibitem[{{Maercker} {et~al.}(2008){Maercker}, {Sch{\"o}ier}, {Olofsson},
  {Bergman}, \& {Ramstedt}}]{2008A&A...479..779M}
{Maercker}, M., {Sch{\"o}ier}, F.~L., {Olofsson}, H., {Bergman}, P., \&
  {Ramstedt}, S. 2008, \aap, 479, 779

\bibitem[{{Mangum} \& {Wootten}(1993)}]{1993ApJS...89..123M}
{Mangum}, J.~G. \& {Wootten}, A. 1993, \apjs, 89, 123

\bibitem[{{Maret} {et~al.}(2004){Maret}, {Ceccarelli}, {Caux}, {Tielens},
  {J{\o}rgensen}, {van Dishoeck}, {Bacmann}, {Castets}, {Lefloch}, {Loinard},
  {Parise}, \& {Sch{\"o}ier}}]{2004A&A...416..577M}
{Maret}, S., {Ceccarelli}, C., {Caux}, E., {et~al.} 2004, \aap, 416, 577

\bibitem[{{Motte} {et~al.}(1998){Motte}, {Andr{\'e}}, \&
  {Neri}}]{1998A&A...336..150M}
{Motte}, F., {Andr{\'e}}, P., \& {Neri}, R. 1998, \aap, 336, 150

\bibitem[{{Muders} {et~al.}(2006){Muders}, {Hafok}, {Wyrowski}, {Polehampton},
  {Belloche}, {K{\"o}nig}, {Schaaf}, {Schuller}, {Hatchell}, \& {van der
  Tak}}]{2006A&A...454L..25M}
{Muders}, D., {Hafok}, H., {Wyrowski}, F., {et~al.} 2006, \aap, 454, L25

\bibitem[{{M{\"u}ller} {et~al.}(2005){M{\"u}ller}, {Schl{\"o}der}, {Stutzki},
  \& {Winnewisser}}]{2005JMoSt.742..215M}
{M{\"u}ller}, H.~S.~P., {Schl{\"o}der}, F., {Stutzki}, J., \& {Winnewisser}, G.
  2005, Journal of Molecular Structure, 742, 215

\bibitem[{{M{\"u}ller} {et~al.}(2001){M{\"u}ller}, {Thorwirth}, {Roth}, \&
  {Winnewisser}}]{2001A&A...370L..49M}
{M{\"u}ller}, H.~S.~P., {Thorwirth}, S., {Roth}, D.~A., \& {Winnewisser}, G.
  2001, \aap, 370, L49

\bibitem[{{Nagaoka} {et~al.}(2005){Nagaoka}, {Watanabe}, \&
  {Kouchi}}]{2005ApJ...624L..29N}
{Nagaoka}, A., {Watanabe}, N., \& {Kouchi}, A. 2005, \apjl, 624, L29

\bibitem[{{Nummelin} {et~al.}(2000){Nummelin}, {Bergman}, {Hjalmarson},
  {Friberg}, {Irvine}, {Millar}, {Ohishi}, \& {Saito}}]{2000ApJS..128..213N}
{Nummelin}, A., {Bergman}, P., {Hjalmarson}, {\AA}., {et~al.} 2000, \apjs, 128,
  213

\bibitem[{{Pardo} {et~al.}(2001){Pardo}, {Cernicharo}, \&
  {Serabyn}}]{2001ITAP...49.1683P}
{Pardo}, J.~R., {Cernicharo}, J., \& {Serabyn}, E. 2001, IEEE Transactions on
  Antennas and Propagation, 49, 1683

\bibitem[{{Parise} {et~al.}(2004){Parise}, {Castets}, {Herbst}, {Caux},
  {Ceccarelli}, {Mukhopadhyay}, \& {Tielens}}]{2004A&A...416..159P}
{Parise}, B., {Castets}, A., {Herbst}, E., {et~al.} 2004, \aap, 416, 159

\bibitem[{{Parise} {et~al.}(2006){Parise}, {Ceccarelli}, {Tielens}, {Castets},
  {Caux}, {Lefloch}, \& {Maret}}]{2006A&A...453..949P}
{Parise}, B., {Ceccarelli}, C., {Tielens}, A.~G.~G.~M., {et~al.} 2006, \aap,
  453, 949

\bibitem[{{Parise} {et~al.}(2002){Parise}, {Ceccarelli}, {Tielens}, {Herbst},
  {Lefloch}, {Caux}, {Castets}, {Mukhopadhyay}, {Pagani}, \&
  {Loinard}}]{2002A&A...393L..49P}
{Parise}, B., {Ceccarelli}, C., {Tielens}, A.~G.~G.~M., {et~al.} 2002, \aap,
  393, L49

\bibitem[{{Parise} {et~al.}(2009){Parise}, {Leurini}, {Schilke}, {Roueff},
  {Thorwirth}, \& {Lis}}]{2009A&A...508..737P}
{Parise}, B., {Leurini}, S., {Schilke}, P., {et~al.} 2009, \aap, 508, 737

\bibitem[{{Pottage} {et~al.}(2004){Pottage}, {Flower}, \&
  {Davis}}]{2004MNRAS.352...39P}
{Pottage}, J.~T., {Flower}, D.~R., \& {Davis}, S.~L. 2004, \mnras, 352, 39

\bibitem[{{Roberts} \& {Millar}(2007)}]{2007A&A...471..849R}
{Roberts}, H. \& {Millar}, T.~J. 2007, \aap, 471, 849

\bibitem[{{Rodgers} \& {Charnley}(2002)}]{2002P&SS...50.1125R}
{Rodgers}, S.~D. \& {Charnley}, S.~B. 2002, \planss, 50, 1125

\bibitem[{{Roueff} {et~al.}(2005){Roueff}, {Lis}, {van der Tak}, {Gerin}, \&
  {Goldsmith}}]{2005A&A...438..585R}
{Roueff}, E., {Lis}, D.~C., {van der Tak}, F.~F.~S., {Gerin}, M., \&
  {Goldsmith}, P.~F. 2005, \aap, 438, 585

\bibitem[{{Roueff} {et~al.}(2007){Roueff}, {Parise}, \&
  {Herbst}}]{2007A&A...464..245R}
{Roueff}, E., {Parise}, B., \& {Herbst}, E. 2007, \aap, 464, 245

\bibitem[{{Roueff} {et~al.}(2000){Roueff}, {Tin{\'e}}, {Coudert}, {Pineau des
  For{\^e}ts}, {Falgarone}, \& {Gerin}}]{2000A&A...354L..63R}
{Roueff}, E., {Tin{\'e}}, S., {Coudert}, L.~H., {et~al.} 2000, \aap, 354, L63

\bibitem[{{Rybicki} \& {Hummer}(1991)}]{1991A&A...245..171R}
{Rybicki}, G.~B. \& {Hummer}, D.~G. 1991, \aap, 245, 171

\bibitem[{{Rybicki} \& {Hummer}(1992)}]{1992A&A...262..209R}
{Rybicki}, G.~B. \& {Hummer}, D.~G. 1992, \aap, 262, 209

\bibitem[{{Snow} {et~al.}(2008){Snow}, {Destree}, \&
  {Welty}}]{2008ApJ...679..512S}
{Snow}, T.~P., {Destree}, J.~D., \& {Welty}, D.~E. 2008, \apj, 679, 512

\bibitem[{{Turner}(1990)}]{1990ApJ...362L..29T}
{Turner}, B.~E. 1990, \apjl, 362, L29

\bibitem[{{Turner}(2001)}]{2001ApJS..136..579T}
{Turner}, B.~E. 2001, \apjs, 136, 579

\bibitem[{{Vassilev} {et~al.}(2008){Vassilev}, {Meledin}, {Lapkin}, {Belitsky},
  {Nystr{\"o}m}, {Henke}, {Pavolotsky}, {Monje}, {Risacher}, {Olberg},
  {Strandberg}, {Sundin}, {Fredrixon}, {Ferm}, {Desmaris}, {Dochev},
  {Pantaleev}, {Bergman}, \& {Olofsson}}]{2008A&A...490.1157V}
{Vassilev}, V., {Meledin}, D., {Lapkin}, I., {et~al.} 2008, \aap, 490, 1157

\bibitem[{{Ward-Thompson} {et~al.}(1989){Ward-Thompson}, {Robson}, {Whittet},
  {Gordon}, {Walther}, \& {Duncan}}]{1989MNRAS.241..119W}
{Ward-Thompson}, D., {Robson}, E.~I., {Whittet}, D.~C.~B., {et~al.} 1989,
  \mnras, 241, 119

\bibitem[{{Watanabe} \& {Kouchi}(2008)}]{2008PrSS...83..439W}
{Watanabe}, N. \& {Kouchi}, A. 2008, Progress In Surface Science, 83, 439

\bibitem[{{Wouterloot} {et~al.}(2005){Wouterloot}, {Brand}, \&
  {Henkel}}]{2005A&A...430..549W}
{Wouterloot}, J.~G.~A., {Brand}, J., \& {Henkel}, C. 2005, \aap, 430, 549

\end{thebibliography}
\end{document}